\documentclass[final,3p,times]{elsarticle}

\usepackage{amssymb}
\usepackage{amsthm}
\usepackage{url}
\usepackage{amsmath}
\usepackage{cuted}
\usepackage{color}  
\usepackage{lineno}

\journal{Physica A: Statistical Mechanics and its Applications}

\begin{document}
\bibliographystyle{elsarticle-num}
\begin{frontmatter}



\title{Exploring the foundation of social diversity and coherence with a novel attraction-repulsion model framework}

\author{Peng-Bi Cui\fnref{label1}}
\ead{cuisir610@gmail.com}
\affiliation[label1]{International Academic Center of Complex Systems, Beijing Normal University, Zhuhai 519087, China}

\begin{abstract}
Opinion evolution is generally subject to global neutral consensus, fragmentation state or polarization. Additionally, there is one widely-existed state --``harmony with diversity" in which individuals freely express various viewpoints to sustain integration of social diversity, but at the same time shared values ensure social coherence. Such state can thus be considered as the foundation of social diversity and coherence, however, which has never attracted research attention. Its formation mechanism still remains unclear. To address this issue, this study proposes an attraction-repulsion model based on the general simple assumption that individuals tend to either reach an agreement with shared opinions or to amplify difference from others with distant opinions. It allows us to take account into the three core parameters: interaction strength, individuals' susceptibility and tolerance to others' opinions. We are concerned with the effect of not only time-varying topology but also fixed interactions imposed by static social network, where the tasks of heterogeneous individuals' attributes are also performed. Remarkably, the simple model rules successfully generate the three above  phases except for fragmentation, along with three different transitions and the triple points. We find that sufficient susceptibility, intermediate interaction strength and high tolerance can benefit a balance between repulsive and attractive forces, and thus the emergence of "harmony with diversity". However, fixed interactions can introduce cluster-level self-reinforced mechanism which can unexpectedly promote polarization. Heterogeneous susceptibility or tolerance turns out to be an inhibiting factor, which should be avoided. A method to identify the phase boundaries through computing the maximum susceptibility of entropy and stand deviation of opinions, confirmed by numerical simulations, allows us to build phase diagrams and to locate where the triple points are. For the first time, this study focuses on the formation of ``harmony with diversity", and the findings provide profound insights into the foundation of societal diversity and coherence.
\end{abstract}



\end{frontmatter}


\section{Introduction}
\label{sec:introduction}
Opinion evolution is about crucial issue which may generally rises three distinct states: 
(i) Global consensus (GC state) in which people's view evolves towards one common opinion, showing a narrow unimodal distribution centering on the neutral consensus~\cite{Axelrod1997,Huang2023}. 
(ii) Fragmentation (F state) is the state with more than two distinct clusters in the opinion space~\cite{Mas2010,Noorazar2020}.
(iii) Polarization (P state) into two opposing clusters or camps which is characterized by a bimodal distribution around the neutral consensus~\cite{Mark2003,Baldassarri2007,Castellano2009,Guilbeault2018,Allcott2020}, further making compromise impossible and giving rise to social opposition~\cite{Guilbeault2018,Allcott2020}.  
Especially, polarization is here conceived as a process of growing constraint in people's viewpoints, and further emergence of alignments along multiple or even opposite lines of potential disagreement with respect to political views, immigration, biotechnology applications, LGBT rights, mask wearing and viral nucleic acid detection against COVID-19~\cite{Axelrod1997,Galam2002,Galam20071,Galam20072,Guilbeault2018,Allcott2020,Galam2020}.

In addition, a striking state besides GC, F and P states cannot be neglected, in which individuals' views or positions self-organize into only one cluster with wide spectrum (i.e., occupied opinion space) and fluctuations. 
It is definitely different from F and GC states that, with respect to ideological issues, the emergence of one opinion cluster with wide spectrum can be defined as a desired state of ``harmony with diversity" (HD state), which has also been stressed by Confucius. 
Since the presence of HD state reveals that individuals can freely express various viewpoints while the free internal debates result in fluctuations, and at the same time shared values ensure social coherence and the avoidance of ideological split, or even sharp conflicts. 
This robust empirical phenomenon is actually vary crucial to healthy functioning of social systems, and therefore regarded as the foundation of societal diversity and coherence~\cite{Renn2022}.
For example, except the United Kingdom, member states of the European Union (EU) over the last two decades have been straining willingness to comply with EU agreements, aiming to sustain an internal single market based on the common Motto ``United in Diversity", regardless of the fact they have various economic demands~\cite{Schimmelfennig2015}. 
Moreover, the softness of the laws or decrees targeting COVID-19, if they are embodied in a basic loose national agreement that this disease is harmful and infectious, tends to guarantee least controversial measures such as wearing face masks, social distancing and home quarantine~\cite{Milosh2020,Tan2020}. 
Unlike shelter-in-place ordinances, lockdowns, or capacity restrictions, the three measures can diminish the spread of an infection, and simultaneously do least harm to economic activity~\cite{Goethals2020,Milosh2020,Mitze2020,Fischer2020,Chu2020,Perra2021}. 
However, to the best of our knowledge, the mechanism behind  the emergence of HD state still remain unclear.

Considerable efforts have been recently put into modeling polarization dynamics to uncover the mechanisms behind it~\cite{Castellano2009,Kossinets2009,Vasconcelos2019,De2020,Baumann2020,Stewart2021,Santos2021,Axelrod2021,Chu2021,Baumann2021,Jusup2022}. One of progresses is to measure polarization with the variance or SD of continuous ideological positions in a population~\cite{Axelrod2021,Santos2021,Macy2021}. 
However, to the best of our knowledge, the emergence and persistence of HD state has never attracted research attention. 
Since most computational attempts are trapped in the framework of two-state model, failing to additionally reproduce a HD state. 
The main approaches of these studies are based on traditional attraction-repulsion model (ARM) with a confidence bound defining a reference that an individual compares with the opinion distance to a neighbor, and based on it, the individuals decide to move closer or farther from its neighbor~\cite{Mark2003,Baldassarri2007,Santos2021,Axelrod2021,Macy2021,Schelling1971,Taber2006,Acemoglu2011,Flache2011,Bail2018}. 
According to such mechanism, the contest between attracted and repelled forces can accordingly facilitate either GC or P~\cite{Mark2003,Baldassarri2007}. 
Alternatively, some modeling approaches investigate the coevolution of the network topology and opinion dynamics as the result of reinforcement and rewiring activities driven by structural similarity~\cite{Santos2021} or homophily~\cite{Baumann2020,Baumann2021}: the two individuals sharing similar connections or opinions are more likely to interact. 
Whatever the case, the HD state does not appear to emerge. 
Although, some seminal works on cultural diversity show that the emergence of clustering phase is possible~\cite{Mas2010}, GC and P states cannot be theoretically reproduced within the same model rules. 
Moreover, due to the lack of HD state, the physical pictures of opinion evolution from the attempts capturing biased assimilation~\cite{Dandekar2013} or information accumulation~\cite{Flache2017} are not integrated. 
The problem rises two fundamental questions we want to answer in this study: whether simple settings can stably generate HD state in addition to GC and P states, and how to effectively distinguish the three phases from each other. 

Influence plays a decisive role in driving the evolution of individuals' opinions, which is responsible for individuals' tendency to either reach agreement with shared opinions (positive influence)~\cite{Axelrod1997,Deffuant2000,Hegselmann2002,Castellano2009} or to amplify difference from others with distant opinions (negative influence)~\cite{Mark2003,Baldassarri2007,Schelling1971,Taber2006,Flache2011,Bail2018,Chen2017}.  
Consequently, the attraction-repulsion model (ARM) can well curve opinion dynamics by setting a threshold for tolerance of disagreement. Within the framework, rules specify the mechanisms for interaction between individuals, they can thus provide insight about important mechanisms and the role they play in regulating emergent properties of system, and highlight the consequences from a few simple assumptions. 
In addition, other mechanisms such as homophily and media influence can be more easily incorporated into ARM to successfully map more empirical polarization changes~\cite{Gargiulo2016,Lu2019}.

Therefore, to address the above two questions, we develop a novel agent-based attraction-repulsion model (ABARM) featuring the emergence of GC, HD and P states from microscopic interactions between individuals. 
Therefore, this paper focuses narrowly on the phase analysis with respect to ideological issues, rather than elucidating new mechanisms behind polarization. 
In our paradigm, opinion evolution rules are built on the general simple assumption that individual's attraction to or repulsion from others is only dominated by their opinion similarity in one-dimensional topic space. 
It allows us to uncover the exact mechanisms that yield particular outcomes, without loss of consideration of three core parameters: interaction strength, individuals' susceptibility and tolerance to different opinions of others. 
We are concerned with the effect of not only time-varying topology but also the static social network (part of Facebook friendship network) on opinion dynamics. 
Moreover, the tasks with heterogeneous susceptibility or tolerance are also considered and performed to make the study closer to the reality.

Regardless of the interaction structures and the heterogeneous attributes, our novel model successfully generates the three states: (i) global convergence towards a neutral consensus, (ii) a fluctuated cluster with a wide spectrum of opinion, (iii) polarization which opinions split into two opposing camps, which are defined as GC, HD and P states, respectively. 
It can help us uncover the mechanism for facilitating HD state. In addition to the standard deviation (SD) of population opinion, we creatively define opinion entropy as a novel order parameter of the system. 
Crucially, a novel available method which indicating the maximum susceptibility of opinion entropy is correspondingly developed to numerically identify the boundaries between the phases. 
Note the good agreement between simulations and the estimated boundaries, validating the phase-identification method. 
We thus conclude that the three phases are distinct in our model. 
Interestingly, comparison of the results from time-varying networks and part of Facebook friendship network reveals that the reinforcement from spatial opinion clusters can increase the likelihood of high-level polarization, instead of extreme polarization. 
This highlights complex roles of group-level struggles. 
In addition, the present model successfully reproduces the fluctuated behavior of opinion cluster in spectrum space without the noise effect, which has been previously considered as the central mechanism to facilitate the fluctuation. 
For the first time, our study proposes a basic framework to explore the formation of HD state, and thus the unique dynamic features of HD. 

The rest of the paper is organized as follows. 
In Sec.~\ref{sec:model_method}, we introduce the model, as well as the phase-identification method. 
In Sec.~\ref{subsec:wellmixed}, we firstly simulate our model with homogeneous and heterogeneous individual attributes on time-varying networks, and then on parts of empirical social networks with fixed interactions in Sec.~\ref{subsec:network}, attempting to make deep comparison. 
We conclude with a summary of the results and an outlook for future studies in Sec.~\ref{sec:conclusion}.

\section{Model and Phase-Identification Method}
\label{sec:model_method}
\subsection{Model}
\label{subsec:model}
The developed ABARM basically assumes a threshold of tolerance to determine whether a repulsive or attractive interaction happens, in line with the assumption of bounded confidence which has been widely confirmed by previous studies~\cite{Deffuant2002,Hegselmann2002,Axelrod2021}. The threshold defines a reference distance in terms of opinion difference, based on which an agent would decides whether to move closer or farther from the selected neighbor. If opinion difference between the individual and its neighbor is smaller than the threshold, an individual will tolerate the position of its neighbor, and move closer to the neighbor. Otherwise, it will move farther from the neighbor. After defining the original model, in turn, we can proceed our explorations by accomplishing the following. (i) Make the individuals' attributes in response to the others' position heterogeneous. (ii) Next, endow the parts of social networks with local structure to fix the interactions among individuals. A systematic comparison of the three cases (including the basic model) with regard to phase areas allows us to fully understand the formation of HD.   

The basic model considers a population of size $N$ where each individual is typified by an opinion $x_{i}(t)$ at time $t$, which is a real number in the interval $x_{i}(t)\in[-10,~+10]$. Inspired by recent studies to capture polarization process on social media networks~\cite{Baumann2020,Santos2021,Baumann2021,Leonard2021}, the updating of individual's opinion $x_{i}(t)$ is solely driven by the interactions among individuals and is described by the following $N$ coupled ordinary differential equations:
\begin{eqnarray}
\hspace{-.5cm} \dot{x}_{i}(t) & = &
\begin{cases}
A\tanh(\alpha_{i}D_{ji}(t))\quad \text{if}~|D_{ji}(t)|<T_{i}; \\ 
A\tanh(\alpha_{i}\sigma(D_{ji}(t))(T_{i}-|D_{ji}(t)|))\quad \text{if}~|D_{ji}(t)|\geq T_{i}. 
\end{cases}
\label{eq:asfunztion}
\end{eqnarray}
$D_{ji}(t)=x_{j}(t)-x_{i}(t)$ is the opinion difference between $i$ and $j$ at time $t$. $T_{i}$ denotes the tolerance threshold of individual $i$. $\alpha_{i}$ can be interpreted as the controversialness of the topic, and thus gauge nonlinearity of interaction or susceptibility of individual. Moreover, $\alpha_{i}$ positively associates with the extent to which individual $i$ is passionate, attentive or sensitive, as a result, susceptible to be socially influenced. It is obvious that $T_{i}$ and $\alpha_{i}$ are largely related to intrinsic preferences of individual $i$. Nonlinear shape of the influence function $tanh(x)$ is controlled by $\alpha$. While $A$ quantifies interaction strength, which is actually the upper bound of opinion shift driven by each interaction, indicating that the influence exerted by individuals on others is capped, in accordance with the experimental findings~\cite{Jayles2017}. $\sigma(D_{ji}(t))$ extracts the sign of $D_{ji}(t)$. Note that in our model the opinion difference is responsible for the evolution of opinions, attempting to capture both the repulsive and attractive forces, which differs from the model settings focusing mainly on reinforcement and homophily~\cite{Baumann2020,Santos2021,Baumann2021,Leonard2021}.   

Firstly, we consider a homogeneous population where individuals have uniform attributes $\alpha_{i}=\alpha$, $T_{i}=T$ and $A_{i}=A$, while the activity of each individual is formulated by activity driven (AD) model~\cite{Liu2014,Perra2012,Moinet2015,Baumann2021}, so that opinion evolution is coupled to an underlying time-varying network. More in detail, $k_{i}(t)$ represents the number of interactions they can have with others within a given time step $t$. It hence gives rise to a temporal network curved by the temporal adjacency matrix $A_{ij}(t)$, where $A_{ij}(t)=1$ if individual $i$ contacts individual $j$, otherwise $A_{ij}(t)=0$. More specifically, individual $i$ connects $k_{i}$ distinct random other individuals, such that $k_{i}=\sum^{N}_{j}A_{ij}(t)$ is satisfied throughout the simulation. Considering the empirical statistics that activities of people are generally heterogeneous~\cite{Perra2012,Moinet2015,Baumann2020}, we assume the interactions are extracted from a power-law distribution $p(k)\sim k^{-\gamma}$.  

Intrinsic preferences toward topics are always proved to be heterogeneous in reality~\cite{Duggins2017,Leon2020}. Therefore, we next assign each individual a susceptibility $\alpha_{i}$ or tolerance threshold $T_{i}$, which is correspondingly selected from the following power-law distributions, respectively. 
\begin{eqnarray}
p(\alpha)\sim \alpha^{-\eta}, \label{eq:hdistribution1}\\
p(T)\sim T^{-\xi}.
\label{eq:hdistribution2}
\end{eqnarray}
$\eta$ is susceptibility exponent, and $\xi$ is tolerance threshold exponent. In such case, we still perform simulations on time-varying networks.

Online social networks are increasingly used to access opinion information, engage with COVID-19 vaccines, gun-control, abortion and so on. 
These platforms can reduce barriers and cost to information and, further, allow individuals to freely voice their viewpoints, consequently improving rate of opinion exchanges. 
On the other hand, the structures of social networks combined with other mechanism such as reinforcement can drive the emergence of echo chambers~\cite{Del2016,Garimella2018,Santos2021}, where the segregation in opinion space is reflected in interactions among individuals~\cite{Bail2018}.
Therefore, we finally fix the interactions among individuals by embedding the population into the parts of the social media networks such as Facebook. In such cases, $k_{i}$ corresponds to the number of edges that individual $i$ stretch to its neighbors, and the individual is represented by a node in the networks. As a result, the connected neighbors of each individual keep unchanged. We perform the above two cases on the static networks.
 
In numerical simulations, we set the simulation parameters to be the following values: $N=1000$ for time-varying networks and $\gamma=2.1$, that will be specified when needed. The control parameters of the present model are $A$, $\alpha_{i}$, $T_{i}$, $\eta$ and $\xi$. The final results are obtained from $N_{r}=100$ independent realizations, after at least $500$ time steps. For each simulation realization, the initial opinions attributed to each individual is independently and randomly sampled from the interval $[-1.0,~1.0]$. Then at each time step $t$, opinion evolves as following: (i) In random order, each individual ($i$) is selected from the population. (ii) Within the framework of AD, the initial temporal adjacency matrix $A_{ij}(t)$ is the zero matrix, and $i$ can randomly choose $k_{i}$ new neighbors out of all individuals. Nevertheless, in the cases considering a static social media network, the neighbors of $i$ keep unchanged. (iii) Then, one by one, $i$ compares its opinion with that of the neighbors, attempting to update its opinion according to Eq.~\ref{eq:asfunztion}. 

We operationalize the degree of polarization through the SD in opinions $SD(x_{0} ,..., x_{N})$, and measure the opinion diversity by calculating the opinion entropy ($S$) of the population: $S =\sum^{x_{max}}_{x_{min}}x\rho_{x}$. $\rho_{x}=\frac{N_{x}}{N}$ is the density of individual owning opinion $x$; $N_{x}$ denotes the population of opinion $x$; $x_{min}=-10$ and $x_{max}=10$. 

\subsection{Phase-Identification Method}
\label{sec:method}
In our model, $SD$ gets larger due to increasing degree of global polarization with a bimodal distribution, and the minimum of the polarization $SD=0$ suggests a narrow unimodal distribution and thus the existence of GC state. While HD state corresponds to that an opinion cluster with wide spectrum. The changes of opinion entropy are therefore sensitive to the transitions between different state. Inspired by this fact, we employ the susceptibility of $S$ to numerically determine the thresholds between different sates.
\begin{eqnarray}
\chi(S) & = & \frac{\sqrt{\langle S^{2}\rangle-\langle S\rangle^{2}}}{\langle S\rangle}, \label{eq:varianceS}
\end{eqnarray}
where $\langle S\rangle$ is the ensemble average of $S$, which can be obtained by averaging $S$ from $N_{r}$ independent realizations. $\langle S^{2}\rangle$ is the secondary moment of the ensemble distribution. Based on the principle that $\chi(S)$ exhibits a peak value at the threshold, one can further identify the boundaries between GC, HD and P states.

\section{Results}
\label{sec:results}

\subsection{The results from time-varying networks}
\label{subsec:wellmixed}

\begin{figure*}
\centering
\includegraphics[width=\linewidth]{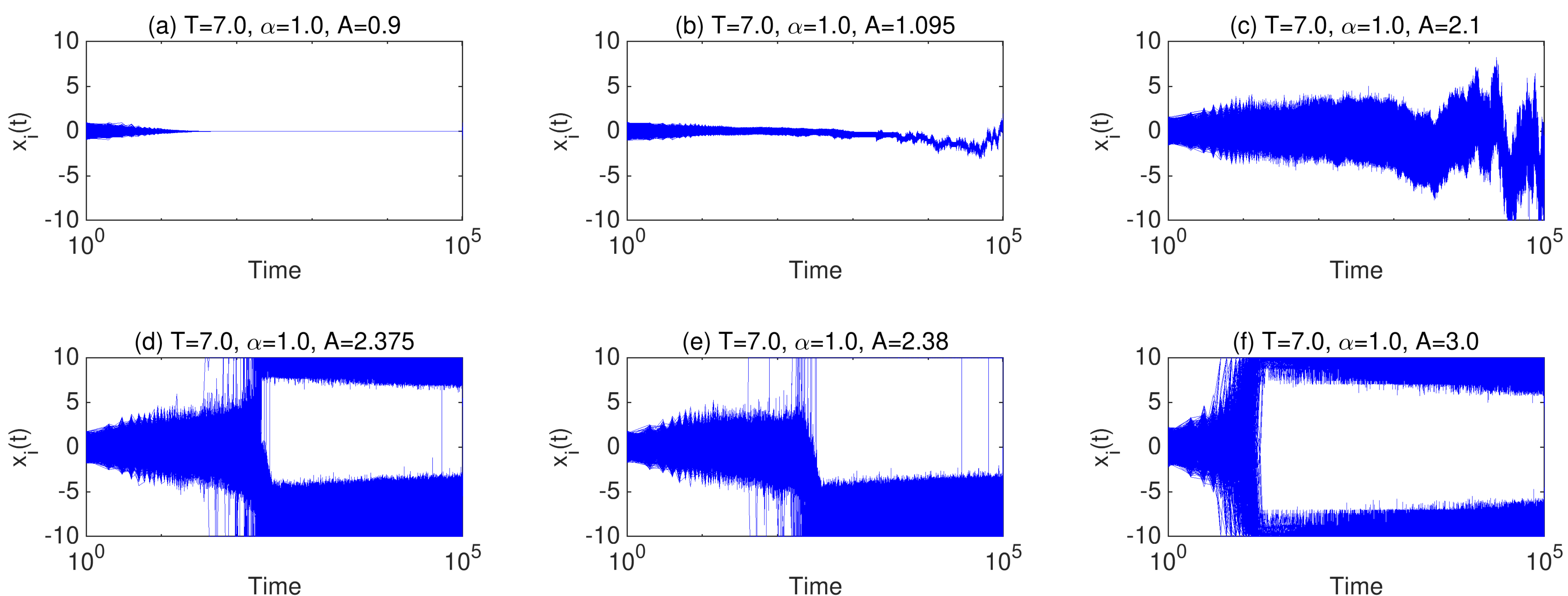}
\caption{Temporal evolution of the individuals' opinions. The six cases are also indicated by the dots in Fig.~\ref{fig:figure3}(a3) for a more explicit presentation. The values of the parameters are listed in the titles of the subplots. 
}
\label{fig:figure1}
\end{figure*}

\begin{figure*}
\centering
\includegraphics[width=\linewidth]{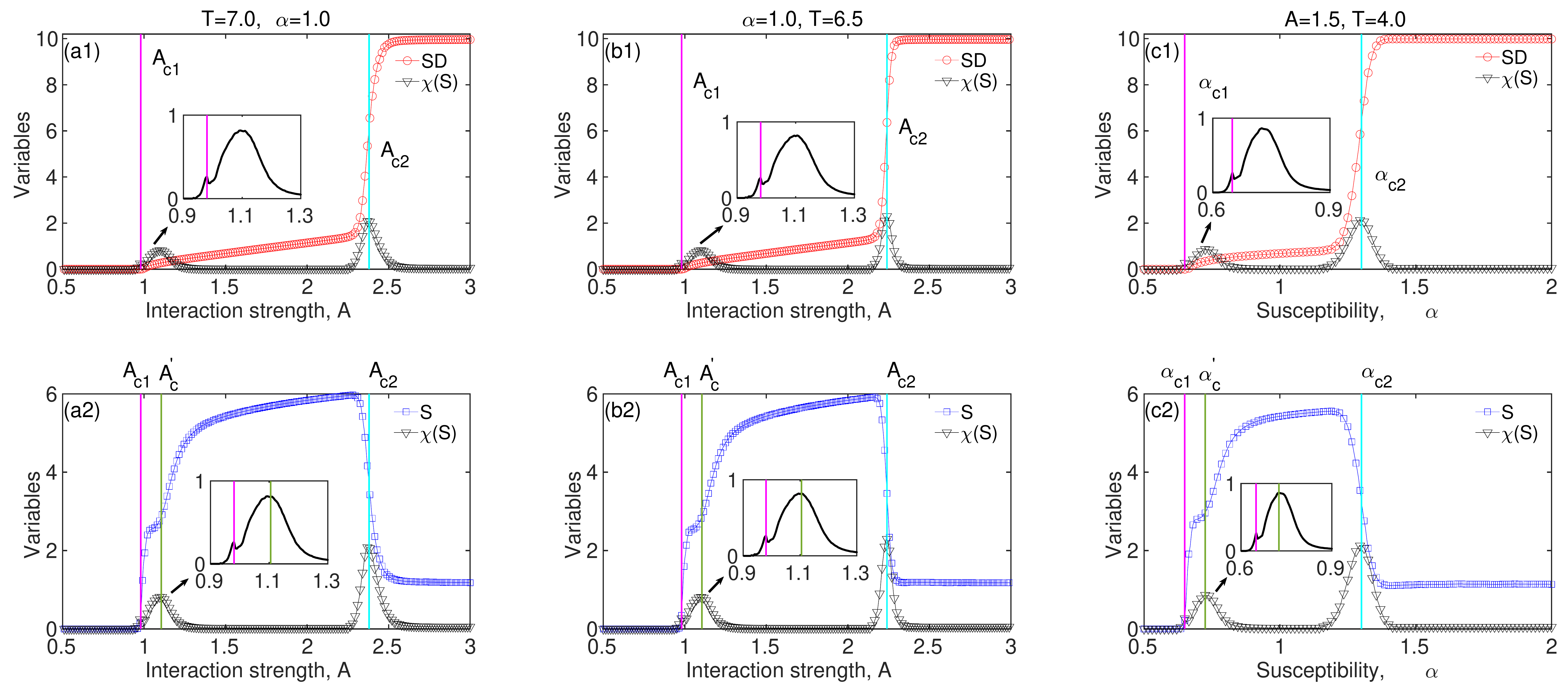}
\caption{The dependence of $SD$ (red circles), $S$ (blue squares), $\chi(S)$ (black cures in the insets) on $A$ or $\alpha$ in time-varying networks. Pink vertical line labels the position of $A_{c1}$ at which HD state begins to appear, while light blue vertical line indicates the position of $A_{c2}$ at which HD state starts to vanish. In particular, $A^{'}_{c}$ denotes an inflection point in $S$ followed by a sharp growth of $S$. The values of other parameters are correspondingly listed in titles of (a1), (b1) and (c1). 
}
\label{fig:figure2}
\end{figure*}

The evolution of a system following Eq.~\ref{eq:asfunztion} is reported in Fig.~\ref{fig:figure1}. 
Three different cases occur in order with increasing interaction strength $A$ or individuals' susceptibility $\alpha$, as shown in Fig.~\ref{fig:figure2}. The population firstly converges to the neutral consensus within a short time (see Fig.~\ref{fig:figure1}(a)) when individuals are more or less independent due to weak interaction.
Then a striking phenomenon occurs at intermediate levels, which produces a mediate polarization that begins to increase slowly after a transition  whose threshold is denoted by $A_{c1}$ (see the first small peaks illustrated in the subplots of Figs.~\ref{fig:figure2}(a1) and (b1)) and then takes hold, corresponding rightly to a highland of entropy (see Figs.~\ref{fig:figure2}(a2) and (b2)). 
In such case, as shown in Figs.~\ref{fig:figure1}(b) and (c), a fluctuated integral opinion cluster in ideological space persists for the entire simulation. 
Specifically, an inflation point of opinion entropy can also be indicated by the position of the first main peak susceptibility of opinion entropy. While spectrum width of the cluster on both sides of the peak is claimed to be rather different, and the spectrum width tends to extend sharply as $A>A^{'}_{c}$ (compare Figs.~\ref{fig:figure1}(b) and (c)).   
The inflation point is dependent on whether the state of opinion cluster with wide spectrum exists or not. 
Though, the interesting phenomenon is qualitatively different from the dependent transition class uncovered by microcanonical statistical analysis~\cite{Qi2018}.
We can find that the system governed by the intermediate range of interaction strength $A_{c1}<A<A_{c2}$ provides sufficient evidence for the existence of HD state. In such parameter range, a balance between repulsive and attractive forces is allowed in most cases, which is rooted in our model setting that opinion shift is bounded by the parameter $A$. Without the noise effect, the HD opinion cluster remains fluctuated, which implies that the balance is dynamic.  

Yet, the polarization starts to rise uncontrollably at the second threshold $A_{c2}$, at which the transition involves an asymmetric temporal trajectory (see Figs.~\ref{fig:figure1}(d) and (e)), and after which the population goes to extreme i.e., P state due to strong negative influence~\cite{Mark2003,Baldassarri2007,Macy2021}. 
The asymmetric temporal trajectory near $A_{c2}$ suggests an asymmetric polarization even if $A\lesssim A_{c2}$ (see asymmetric opinion distribution illustrated in Fig.~\ref{fig:sfigure0}(a2)), resulting from occasional outperformance of repelled force in few realizations (see Fig.~\ref{fig:sfigure0} and corresponding descriptions for more details). Note that asymmetric polarization has also been the focus of recent studies~\cite{Pierson2020,Leonard2021}. 
In P state, opinion symmetrically splits into two opposite extreme camps, as shown in Fig.~\ref{fig:figure1}(f) and Figs.~\ref{fig:sfigure0}(b1) and (b2). 
Obviously, the three different cases correspond to GC, HD and P states, respectively. 
Moreover, we can see in Figs.~\ref{fig:figure2} (c1) and (c2) the similar dependence of $SD$, $S$ and $\chi(S)$ on individuals' susceptibility $\alpha$.

Fig.~\ref{fig:figure2} shows that estimating the entropy can not only identify the boundaries between different phases but also reflect which phase region corresponds rightly to a HD state. Since the wide spectrum results in a highland of entropy. While the SD of opinions in the population is always used to identify either GC or P state~\cite{Santos2021,Axelrod2021,Macy2021}. The information indicated by intermediate SD is still unclear. Therefore, estimating both SD and entropy is a better choice to completely identify the regions of the three states. It must be stressed that the proposed method actually identifies the nature of evolution outcomes from most realizations, regardless of that asymmetric polarization occurs occasionally (as shown in Fig.~\ref{fig:figure1}(d) and Fig.~\ref{fig:sfigure0}(a2)). Moreover, estimating the entropy can curve  more elaborate behaviors of HD state by showing the inflation point.

\begin{figure*}
\centering
\includegraphics[width=\linewidth]{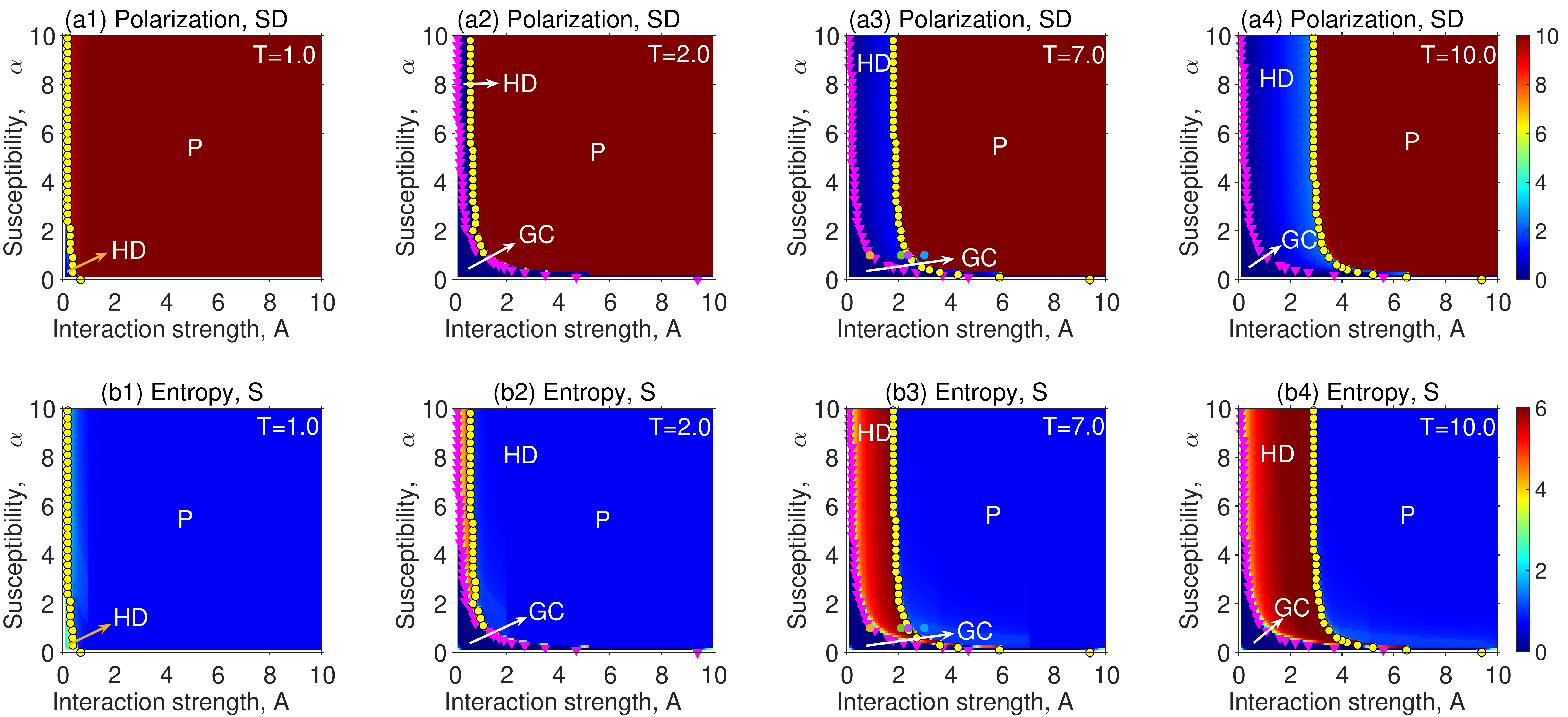}
\caption{Phase diagrams in ($A$,~$\alpha$) space for four different values of $T$. We run simulations with polarization degree $SD$ in (a1)-(a3), and with opinion entropy $S$ in (b1)-(b3). The lines consisting of pink triangles depict the boundaries indicated by the first threshold $A_{c1}$, separating the regions of GC and HD phases; while the lines consisting of yellow circles depict the boundaries indicated by the second threshold $A_{c2}$, separating the regions of HD and P phases, as given by our method. The regions belonging to different states are labeled in the subplots. The six dots correspond to the subplots of six parameter combinations in Fig.~\ref{fig:figure1}.
}
\label{fig:figure3}
\end{figure*}

We then extend the previous observations to a wide range of $(A,~\alpha)$ in Fig.~\ref{fig:figure3}, where the color encodes
the values of $SD$ (top panels) and $S$ (bottom panels), respectively. 
We observe two different transitions: from GC to HD phase when $\alpha$ is small, and from HD to P phase for large $\alpha$. 
The transition from HD to P is largely dependent on $A$ if individuals are sensitive enough to the dissimilarity in opinions ($\alpha$ is not small). 
While the regions of GC are characterized by small values of $A$ and $\alpha$, and regions of HD can be obtained for increasing $A$ and $\alpha$. 
Moreover, a comparison among the top (bottom) panels of Fig.~\ref{fig:figure3} reveals that increasing tolerance threshold $T$ can expand HD regions being robust to change of susceptibility, by simultaneously occupying the regions of GC and P. 
The new interesting phenomenon is largely in accordance with the empirical results~\cite{Spencer2010,Noorazar2020} and the nature of the social moral rules requiring a high level of tolerance, and thus one of novel findings of our present study. 
We can see in Figs.~\ref{fig:figure3}(b1)-(b4) that highlands of opinion entropy perfectly indicate the regions of HD. 
The identified boundaries afford a precise division of the regions of different phases, validating our phase-identification method.

\begin{figure*}
\centering
\includegraphics[width=\linewidth]{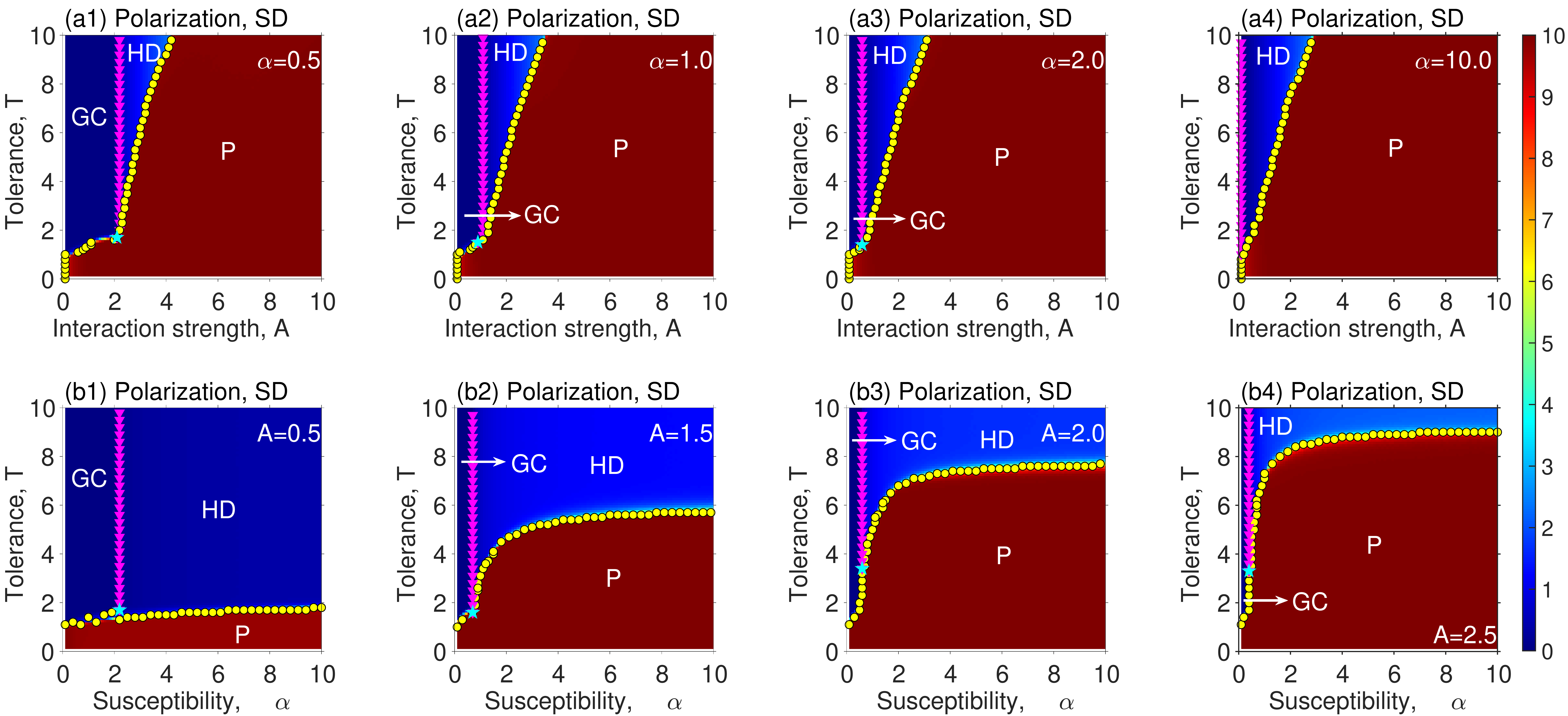}
\caption{(a1)-(a4) Phase diagrams in $(A,~T)$ space for four different values of $\alpha$. (b1)-(b4) Phase diagrams in $(\alpha,~T)$ space for four different values of $A$. We run simulations with polarization degree $SD$ in all subplots. The lines consisting of pink triangles depict the boundaries indicated by the first threshold $A_{c1}$, separating the regions of GC and HD phase; while the lines consisting of yellow circles depict the boundaries indicated by the second threshold $A_{c2}$, separating the regions of HD (GC) and P phases, as given by our method. The regions belonging to different states are labeled in the subplots. The light blue pentagrams indicate the triple points which, however, do not present in $(A,~\alpha)$ space.
}
\label{fig:figure4}
\end{figure*}

Figs.~\ref{fig:figure4}(a1)-(a4) show the phase diagrams in terms of $SD$ in $(A,~T)$ space. 
We find not only transition from GC to P phase if $T$ is small but larger than a certain value, but also transitions from GC to HD phase, and from HD to P phase as $T$ getting larger. 
Therefore, there definitely exists a triple point denoted by a vanishing entropy highland (see the entropy highlands shown in Figs.~\ref{fig:sfigure1}(a1)-(a4)), above which the emergence of GC is completely dominated by $A$, and independent of the change of $T$.
If the tolerance threshold is large enough, weak interaction becomes a fully dominant factor for a global convergence, and intermediate $A$ can generate HD state whose region tends to expand linearly with increasing $T$. 
For all large susceptibility promotes HD to erode the regions of GC, and in turn P occupy the regions of HD, leading to a shrinkage of GC regions until its disappearance when $\alpha=10.0$ (combine with the entropy highlands shown in Figs.~\ref{fig:sfigure1}(a1)-(a4)). 
Figs.~\ref{fig:figure4}(b1)-(b4) offer a comprehensive view of the effects of individual's susceptibility on opinion dynamics for different levels of tolerance. 
We find that, if $\alpha$ is not small any more, polarization is likely to emerge, along with extinction of GC and decreasing likelihood of HD. 
Increasing both tolerance and susceptibility is responsible for HD state. However, in line with the results illustrated in Figs.~\ref{fig:figure2} and \ref{fig:figure3}, more tolerant individuals are required as interaction strengthening (see Figs.~\ref{fig:figure4}(b1)-(b4) and Figs.~\ref{fig:sfigure2}(b1)-(b4)). 
Meanwhile high tolerance would result in saturation effect of large $\alpha$, and role of susceptibility is limited.  
Still, the system can produce a triple point, and there occur three different transitions under certain parameter conditions: from GC to P, GC to HD, and HD to P. 

The results from the basic model on time-varying networks reveal that sufficient susceptibility, intermediate interaction strength and high tolerance are responsible for the dynamic balance between repulsive and attractive forces, further promoting HD state and leading to fluctuations of HD opinion cluster. 
In contrast, strong interaction or low tolerance can make things worse, and the population may get polarized. 
As another novel finding, the system formulated by our model can generally generate triple points in parameter planes of $(A,~T)$ and $(\alpha,~T)$. 

\subsection{The results from heterogeneous attributes}
\label{subsec:heterogeneous}   
Individuals are always heterogeneous in their roles, their power and their capacity to influence others, which is mostly rooted in their heterogeneous intrinsic attributes. Within our model framework, individuals differ in terms of their susceptibility or tolerance. In this subsection, we focus on the effects of heterogeneous susceptibility or tolerance, which follows power law distributions given by Eqs.~(\ref{eq:hdistribution1}) and (\ref{eq:hdistribution2}). This lends a different and distinct perspective to our studies of HD state through comparing the results with the homogeneous case in Subsec.~\ref{subsec:wellmixed}.

\begin{figure*}
\centering
\includegraphics[width=\linewidth]{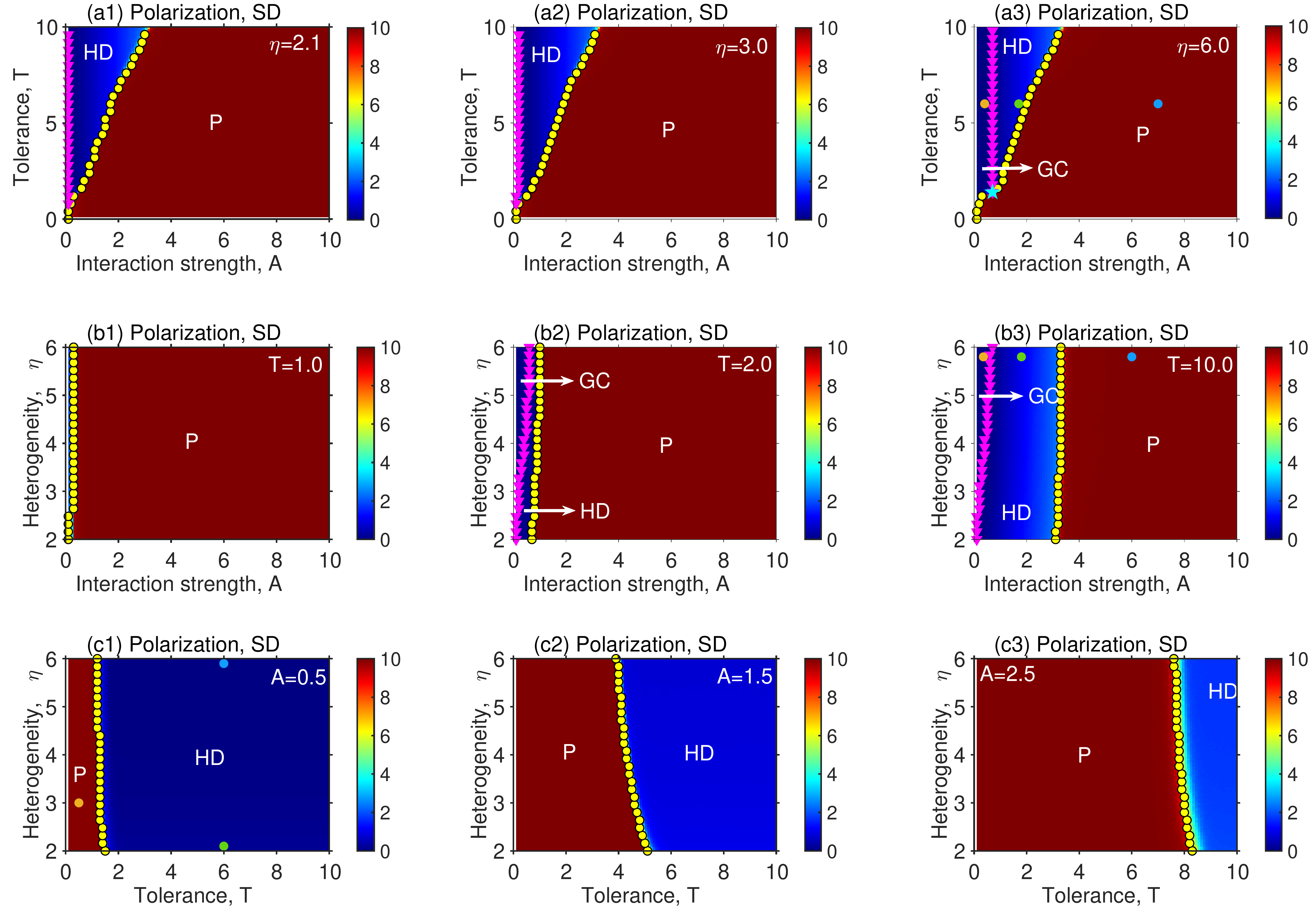}
\caption{Phase diagrams for the population with heterogeneous susceptibility. In more detail, (a1)-(a3) phase diagrams in $(A,~T)$ space for three different values of $\eta$, (b1)-(b3) phase diagrams in $(A,~\eta)$ space for three different values of $T$, (c1)-(c3) phase diagrams in $(T,~\eta)$ for three different values of $A$. We run simulations with polarization degree $SD$ in all subplots. The lines consisting of different markers denote the boundaries between different phases, which are the same as those presented in Fig.~\ref{fig:figure4}. The regions belonging to the three phases are correspondingly labeled in the subplots. The light blue pentagram presented in (a3) indicates the triple point in $(A,~T)$ space. The dots presented in (a3), (b3) and (c1) correspond to the subplots (a1)-(a3), (b1)-(b3) and (c1)-(c3) in Fig.~\ref{fig:figure6}, respectively. 
}
\label{fig:figure5}
\end{figure*}

Fig.~\ref{fig:figure5} depicts phase diagrams in terms of $SD$ in parameter spaces of $(A,~T)$, $(A,~\eta)$ and $(T,\eta)$. 
We firstly note that, in Figs.~\ref{fig:figure5}(a1)-(a3), the layouts of phase diagrams are similar to those illustrated in Figs.~\ref{fig:figure4}(a1)-(a4). 
The difference is that GC emerges only if the susceptibility exponent increases from "scale-free" regime ($\eta<3$) to "small-wold" regime ($\eta>3$) where the value of $SD$ can still identifies three types of transitions, as well as a triple point (see Figs.~\ref{fig:figure5}(a1)-(a3)). 
However, in $(A,~\eta)$ space, things are different. In this case, we find at most two different transitions for high-level tolerance: from GC to HD phase, and from HD to P phase (see Figs.~\ref{fig:figure5}(b1)-(b3)), let alone the existence of the triple point. 
In line with what we have observed in Figs.~\ref{fig:figure5}(a1)-(a3), we can obtain GC state for $\eta\gtrsim 3$ (see Figs.~\ref{fig:figure5}(b1)-(b3)). 
It is evident that $\eta_{c}\approx 3.0$ can thus be considered as a threshold for GC state. 
In addition, intermediate $A$ can generate HD state whose regions expand with $T$, slightly shrink with $\eta$. 
This means that susceptibility heterogeneity prevents the population from converging to GC state. 
As expected, increasing $T$ can effectively promote the population into HD phase, and it becomes harder to achieve the desired state for large $A$ (see Figs.~\ref{fig:figure5}(c1)-(c3) and Figs.~\ref{fig:sfigure2}(c1)-(c3)). 
It is consistent with the observations in homogeneous case. 
Figs.~\ref{fig:figure5}(c1)-(c3) show that HD region gradually expands with $\eta$ along with a shrink of P phase, suggesting a negative role of susceptibility heterogeneity. 
In comparison with the parameters $A$ and $T$, the effect from susceptibility heterogeneity is less pronounced. 

\begin{figure*}
\centering
\includegraphics[width=\linewidth]{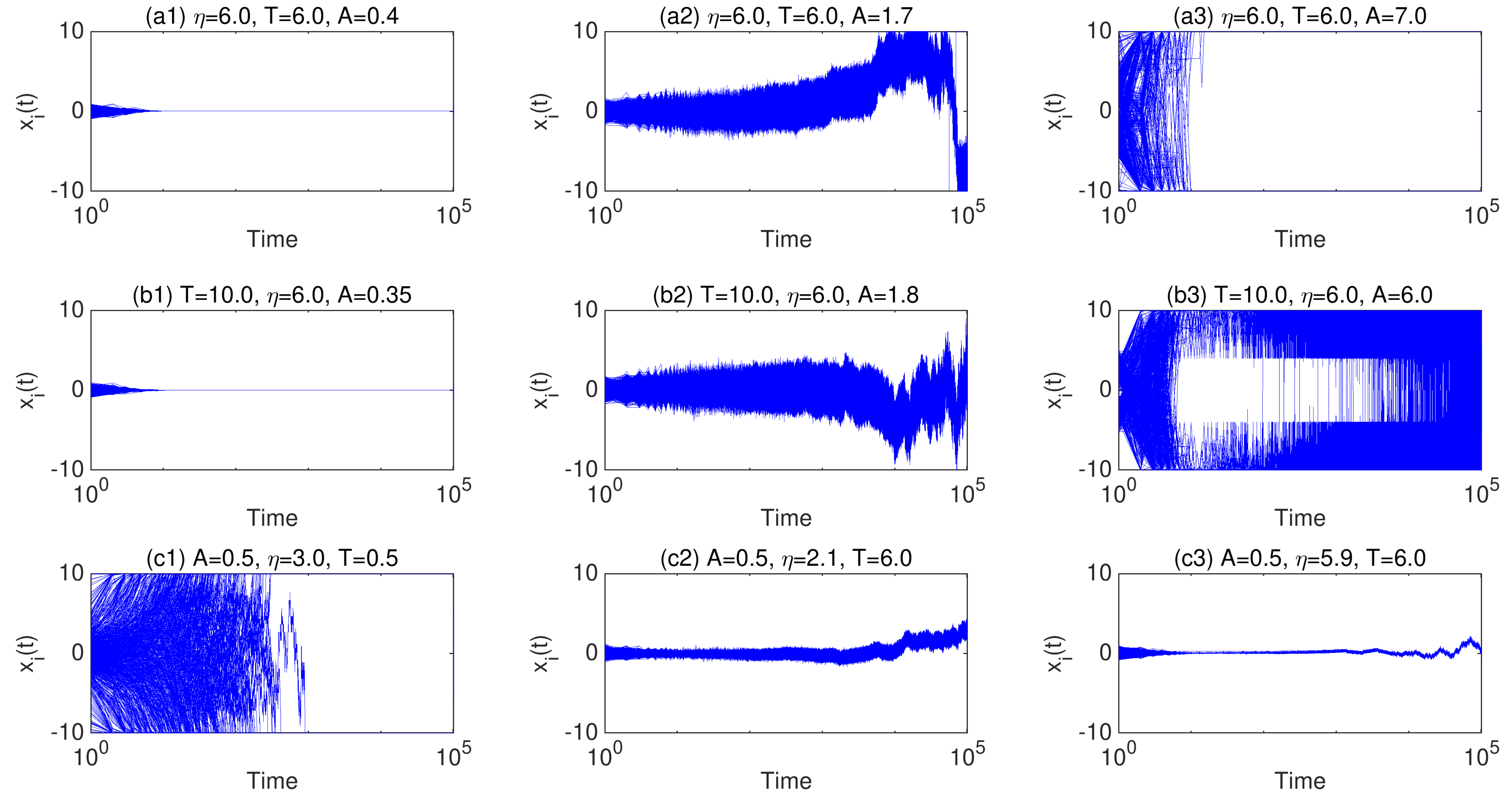}
\caption{Temporal evolution of the individuals' opinions for the population with heterogeneous susceptibility. The values of parameters are correspondingly listed in the titles of the subplots.
}
\label{fig:figure6}
\end{figure*}

Correspondingly, Fig.~\ref{fig:figure6} further approves that the dynamic behaviors do not qualitatively change by exhibiting three states of opinion evolution. 
Especially, the middle subplots, as well as Fig.~\ref{fig:figure6}(c3), show the temporal behaviors of individuals' opinions in HD state, where the opinion cluster is fluctuating irregularly with time, but integral even when the cluster touches the boundary of ideological space. 
This further supports our statement about the system going into HD state where patterns of opinion cluster keep robust against time.  

\begin{figure*}
\centering
\includegraphics[width=\linewidth]{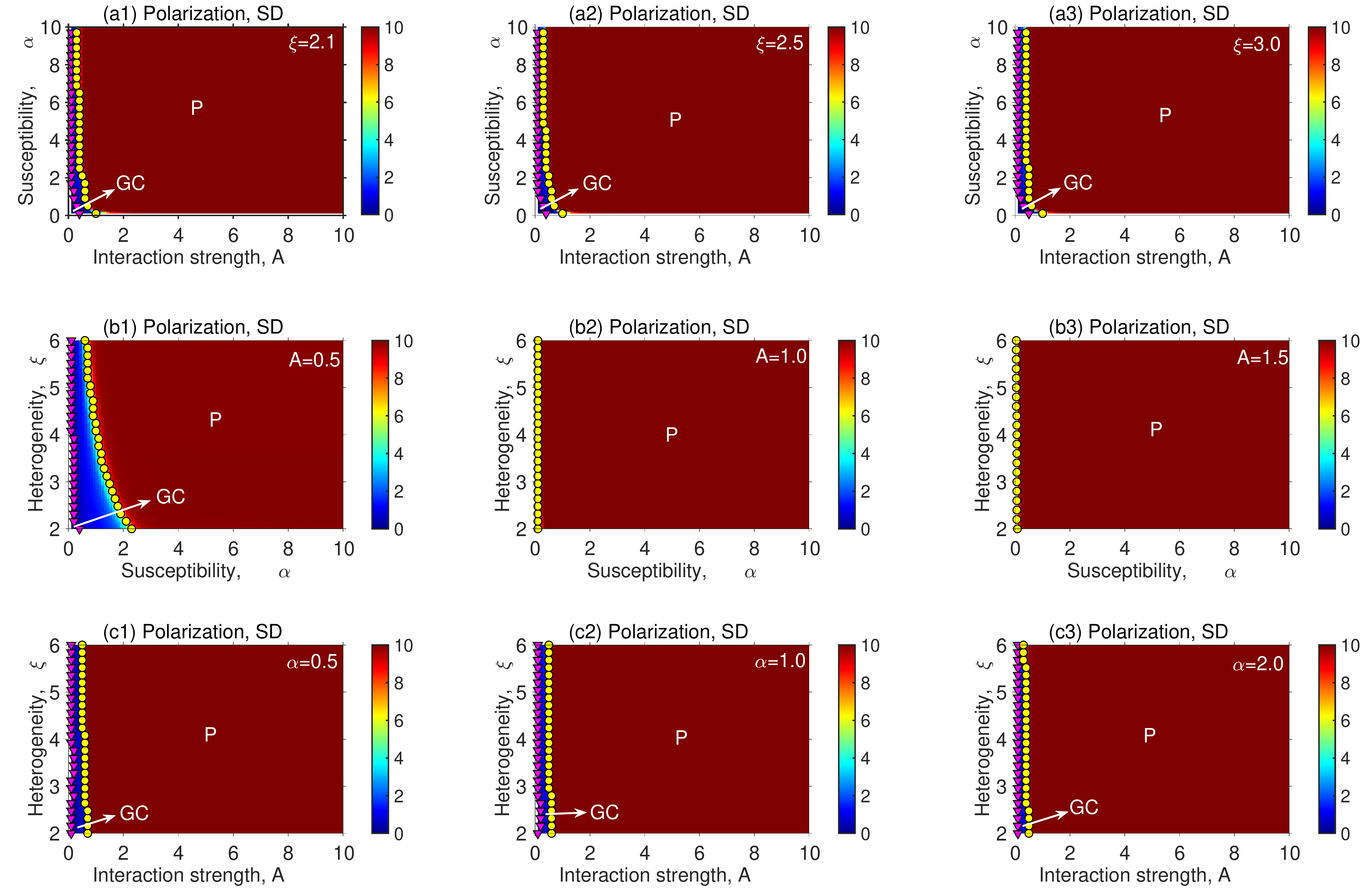}
\caption{Phase diagram for the population with heterogeneous tolerance. In more detail, (a1)-(a3) phase diagrams in $(A,~\alpha)$ space for three different values of $\xi$, (b1)-(b3) phase diagrams in $(\alpha,~\xi)$ space for three different values of $A$, (c1)-(c3) phase diagrams $(A,~\xi)$ for three different values of $\alpha$. The values of parameters are listed in corresponding subplots. We run simulations with $SD$ in all subplots. The lines consisting of either pink triangles or yellow circles depict the boundaries separating the regions of GC and P states, as given by our method. The regions of different phases are correspondingly labeled in the subplots.
}
\label{fig:figure7}
\end{figure*}

We next focus on the effect of heterogeneous individuals' tolerance on opinion dynamics. 
Remarkably, the population with heterogeneous tolerance leads to a qualitative change of the results with respect to both phase diagram and temporal evolution. Tolerance heterogeneity can largely increase the tendency of the population getting trapped in P state. 
Fig.~\ref{fig:figure7} shows that the population will be finally polarized when $A$ or $\alpha$ is beyond a small value. 
This is definitely in contrast to the convergence properties of bounded confidence models~\cite{Lorenz2010}. 
In such case, most individuals are closed-minded, and hence have a stronger tendency to amplify differences from others if their opinions are slightly different.  
Moreover, saying that the blue regions between GC and P phases (see Figs.~\ref{fig:figure7}(a1)-(a3) and (b1)) correspond to HD phase is not appropriate. 
It is instead qualitatively different from the aforementioned HD state. 
Since we can see in Fig.~\ref{fig:sfigure4} that the population is eventually polarized as the opinion cluster touches the boundaries of the ideological space, and gets absorbed. 
Interaction strength $A$ largely determines the population's outcome (see Figs.~\ref{fig:figure7}(c1)-(c3)). 

\subsection{The results from static networks}
\label{subsec:network} 
We subsequently embed our model into the social networks of size $N=43952$, where individuals interact through fixed connections, so as to explore the joint effects of the network topology and influence mechanism on the formation of HD. The detailed structure parameters of the network are also presented in Fig.~\ref{fig:sfigure5_1}.

\begin{figure*}
\centering
\includegraphics[width=\linewidth]{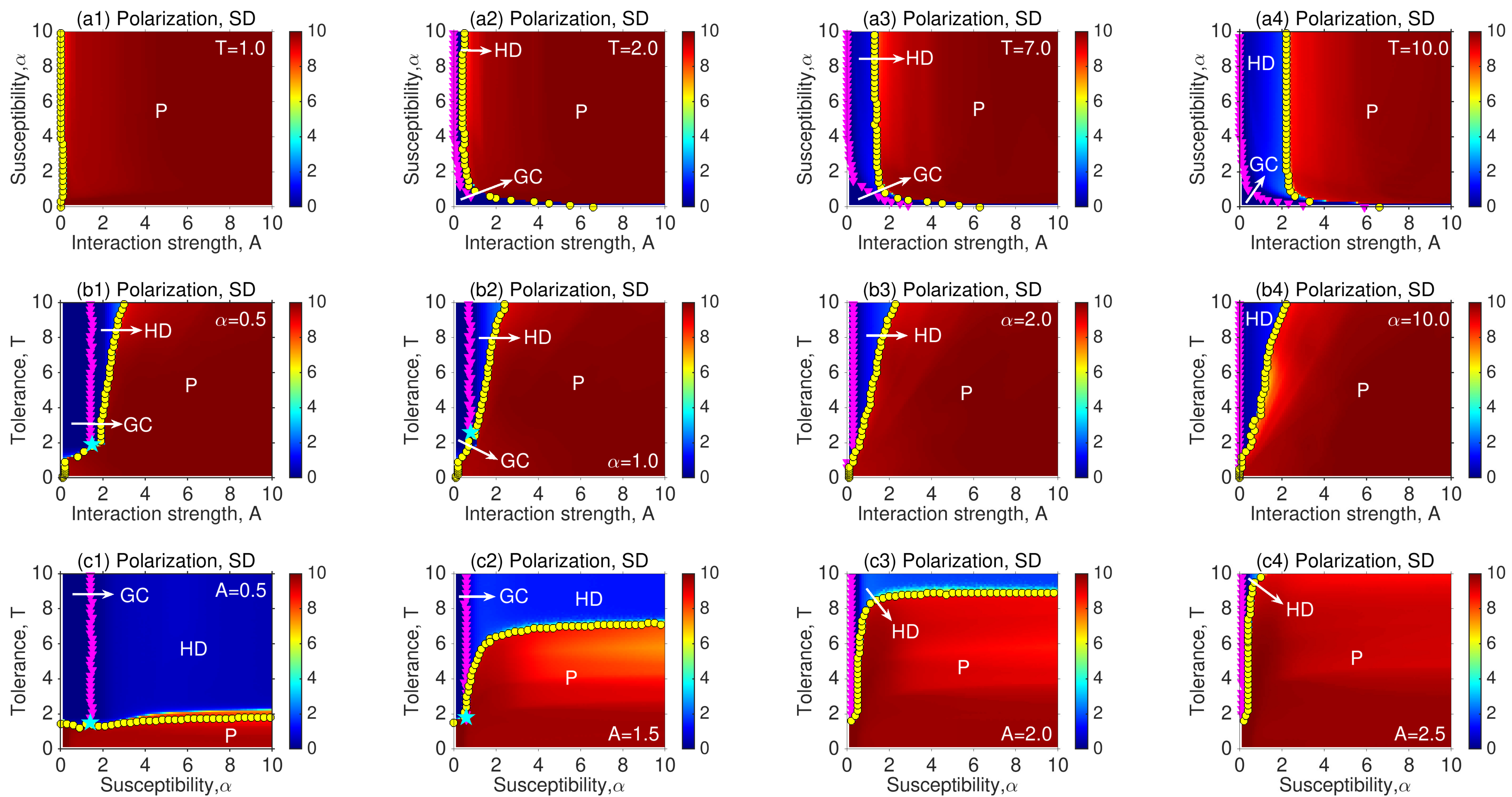}
\caption{Phase diagram for the population embed on part of Facebook network, where interactions among individuals are fixed. (a1)-(a3) Phase diagrams in $(A,~\alpha)$ space for four different values of $T$; (b1)-(b4) Phase diagram in $(A,~T)$ space for four different values of $\alpha$; (c1)-(c4) Phase diagram in $(\alpha,~T)$ space for four different values of $A$. We run simulations with $SD$ in all subplots. The lines consisting of different markers denote the boundaries between different phases, which are the same as those presented in Fig.~\ref{fig:figure4}. The regions of the three phases are correspondingly labeled in the subplots. The light blue pentagrams indicate the triple points which are less frequent. The dots presented in (a3),(b2) and (c1) correspond to the subplots (a1)-(a3), (b1)-(b3) and (c1)-(c3) in Fig.~\ref{fig:sfigure4}, respectively.
}
\label{fig:figure8}
\end{figure*}

The phase diagrams are presented in Fig.~\ref{fig:figure8} in three different parameter spaces: $(A,~\alpha)$, $(A,~T)$ and $(\alpha,T)$. 
Our results more or less remain robust with respect to the engagement of fixed interactions.  
It is evident that the phase layouts are largely dominated by model rules rather than fixed interactions, and thus similar to those illustrated in Figs.~\ref{fig:figure3} and \ref{fig:figure4}. In Fig.~\ref{fig:sfigure5}, we see that the behaviors of $SD$, $S$ and $\chi(S)$ are qualitatively similar to the case of time-varying network (see Fig.~\ref{fig:figure2}). Whereas it is important to note that the introduction of fixed interactions can remarkably give rise to much sharper dependence (see Fig.~\ref{fig:sfigure5}), and much larger regions of P phase where extreme polarization ($SD=10$) coexists with high-level polarization with large $SD$ (see Fig.~\ref{fig:figure8} for the light red regions in P phases). 
Also notice that there are less triple points in comparison with the results reported in Figs.~\ref{fig:figure3} and \ref{fig:figure4}.   

\begin{figure*}
\centering
\includegraphics[width=\linewidth]{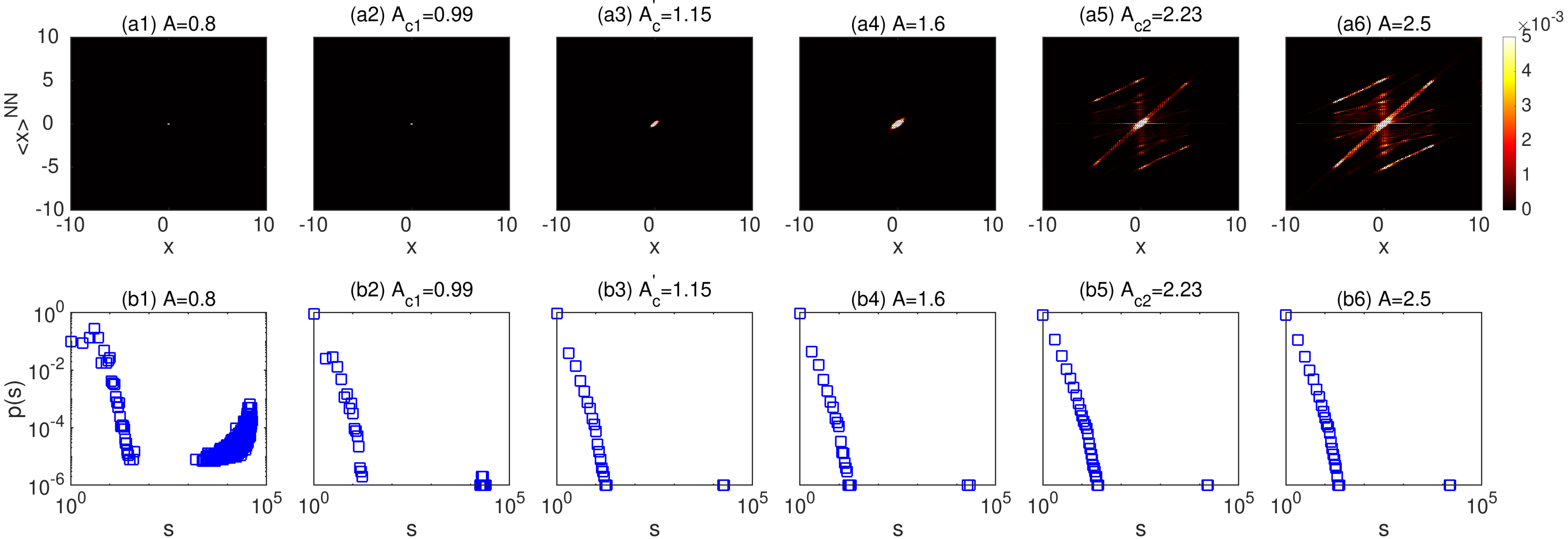}
\caption{(a1)-(a6) Heatmaps for the average opinion of the nearest neighbors $\langle x\rangle^{NN}$ against an individual's opinion $x$, for $N_{r}=500$ simulations of independent realizations. (b1)-(b6) The distributions of positive clusters for six different values of $A$, where positive clusters refer to those consisting of individuals owning positive opinions. The values of other parameters are $T=7.0$ and $\alpha=1.0$.
}
\label{fig:figure9_1}
\end{figure*}

\begin{figure*}
\centering
\includegraphics[width=\linewidth]{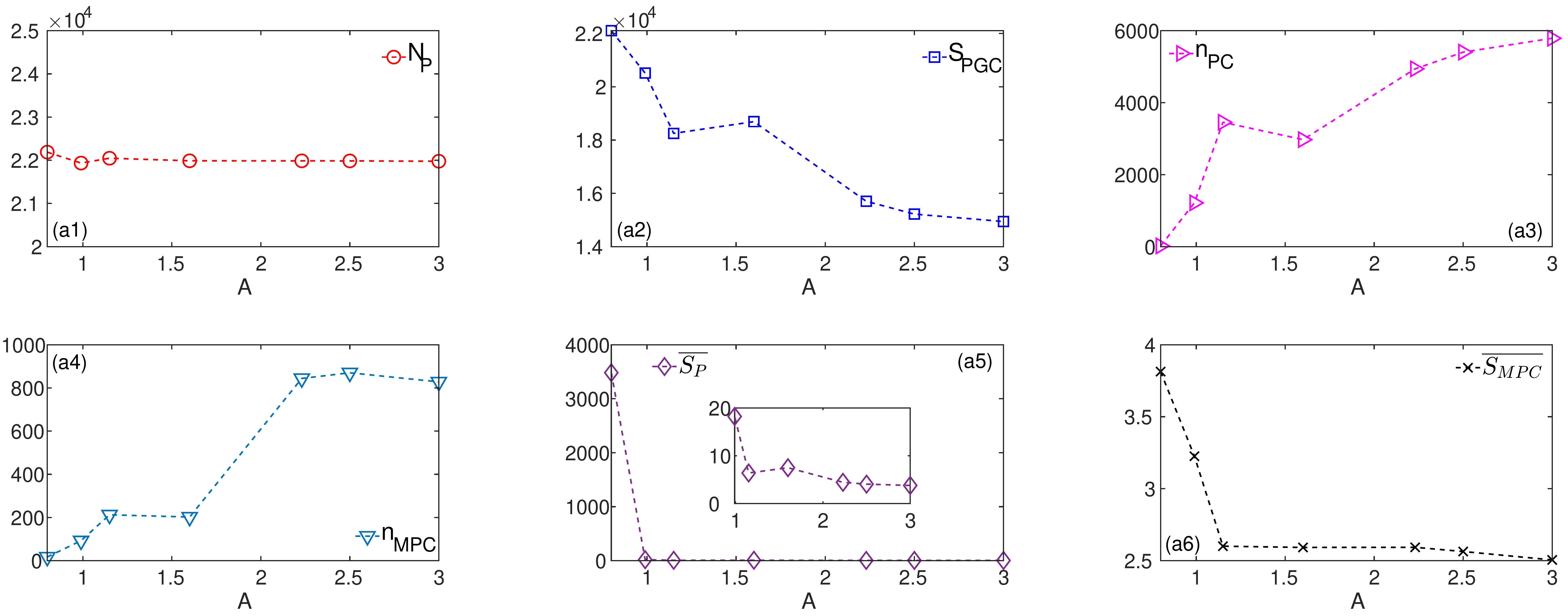}
\caption{(a1) The size of the population with positive opinions $N_{P}$ against $A$; (a2) The size of largest positive cluster (i.e., giant positive cluster) $S_{PGC}$ against $A$; (a3) The number of positive clusters $n_{PC}$ against $A$; (a4) The number of modest positive clusters $n_{MPC}$ against $A$, which exclude positive clusters of size 1 and the largest positive cluster; (a5) The mean size of positive clusters $\overline{S_{P}}$ against $A$, where the subplot presents more clear behavior of $\overline{S_{P}}$ for $A>A_{c1}$; (a6) The mean size of the modest positive clusters $\overline{S_{MPC}}$ against $A$. The six markers presented in (a1)-(a6) rightly correspond the six values of $A$ given in Figs.~\ref{fig:figure9_1}(a1)-(a6) and (b1)-(b6). The values of other parameters are $T=7.0$ and $\alpha=1.0$.
}
\label{fig:figure9_2}
\end{figure*}

\begin{figure*}
\centering
\includegraphics[width=\linewidth]{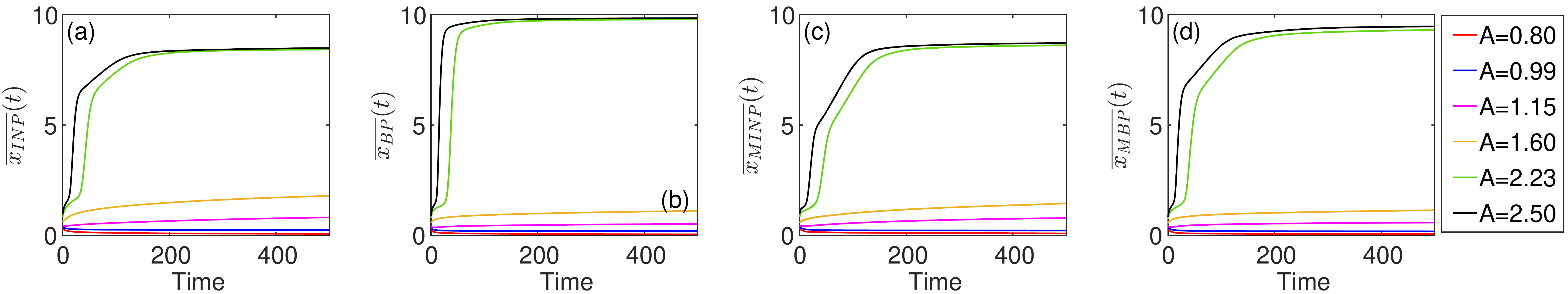}
\caption{(a) Evolution of average opinion of innermost members owning positive opinions $\overline{x_{INP}}(t)$. (b) Evolution of average opinion of individuals at the boundaries of positive clusters $\overline{x_{BP}}(t)$. (c) Evolution of average opinion of innermost members belonging to the modest positive clusters $\overline{x_{MINP}}(t)$. (c) Evolution of average opinion of individuals at the boundaries of modest positive clusters $\overline{x_{MBP}}(t)$. The values of other parameters are $T=7.0$ and $\alpha=1.0$.
}
\label{fig:figure9_3}
\end{figure*}

Figs.~\ref{fig:figure9_1}(a1)-(a6) further illustrate heatmaps of the density of users in the $(x,~\langle x\rangle^{NN})$ plane, for six values of $A$, where the results reveal the correlation between the opinion of an individual $i$, and the average opinion of its nearest neighbors. 
This can measure an individual's preference to connect to peers sharing similar opinions~\cite{Baumann2020}, which fosters information about the effects of cluster-level self-reinforced mechanism. 
In our study, cluster-level self-reinforced mechanism is defined that the like-minded members tend to cluster together and persistently support those on the boundaries to repel distant ones. 
Attributing to the symmetrical nature of our model, the configurations of positive clusters i.e. the clusters of positive opinions, are similar to those of negative clusters with respect to size and number, which is supported by the results reported in Fig.~\ref{fig:figure9_2}(a1) that the summary of positive clusters is approximately equal to half of the whole population i.e. $N_{p}\approx N/2$.  
We thus calculate the statistics of positive clusters by means of statistical variables reported in Figs.~\ref{fig:figure9_1}(b1)-(b6), Fig.~\ref{fig:figure9_2} and Fig.~\ref{fig:figure9_3}, which provide an overall picture of cluster-level self-reinforced mechanism. 
This naturally gives rise to the question whether abundance or size of these spatial opinion clusters guarantee the presence of this mechanism.

As we can see in Figs.~\ref{fig:figure9_1}(b1)-(b6), plotting the distributions of positive clusters for six different values of $A$ shows that there are always giant positive clusters indicated by isolated patterns, as well as power-law cluster-size distributions; which is independent of the values of $A$. Hence the observed phenomenon is not symbol of discontinuous transitions, instead, it is rooted in the model rules and initial assignment of individuals' opinions.

(i) In the parameter range $A<A_{c1}$, the population converges into GC state because the attractive force is dominated when initial opinions are in the narrow range $[-1, 1]$. Whereas the system is divided into two main clusters whose agents have close opinions of different sign, and the boundaries between the clusters are not distinct (see Figs.~\ref{fig:figure9_1}(a1) and (b1), Figs.~\ref{fig:figure9_2}(a2) and (a3), Fig.~\ref{fig:figure9_3}).  

(ii) When $A_{c1}<A<A^{'}_{c}$, a high density of users hold moderate opinions of their own as clusters of size 1, and contest between nodes with opinions of different signs happens frequently to amplify difference from others (compare the first three markers in Figs.~\ref{fig:figure9_2}(a2)-(a5) and see Fig.~\ref{fig:figure9_3}). Therefore, a balance between repulsive and attractive forces is possible, and the population achieves  HD state. At the same time, as Figs.~\ref{fig:figure9_3}(a) and (b) illustrate, $\overline{x_{INP}}(t)$ and $\overline{x_{BP}}(t)$ are small when $A<A^{'}_{c}$. It implies that most individuals are within the largest clusters and moderate. We can thus observe in Fig.~\ref{fig:figure9_1}(a3) that bright spindle-shaped diagonal speckles that stretch the first and third quadrants. 

(iii) For $A^{'}_{c}<A<A_{c2}$, increasing interaction strength makes more individuals split off from the largest cluster to form more clusters of size 1 (compare the third to the fifth markers in Figs.~\ref{fig:figure9_2}(a2), (a3) and (a5)) and more smaller modest clusters (compare the third to the fifth markers in Figs.~\ref{fig:figure9_2}(a3)-(a6)) which are not aligned. Strengthening repulsive force starts to further broaden the spectrum of opinion cluster, giving rise to the sharp growth of $S$ (see Figs.~\ref{fig:sfigure5}(a2) and (b2)). An larger bright spindle-shaped diagonal speckle can thus be observed, as illustrated in  Fig.~\ref{fig:figure9_1}(a4). As $A$ approaching $A_{c2}$, more clusters of size 1 and more smaller modest clusters are generated (see Fig.~\ref{fig:figure9_2}(a2)-(a6)). Therefore, there are longer distinct boundaries between the clusters of different signs, and this results in intra-cluster coherence and intenser cluster-level struggles among clusters of the opposite sides. The individuals at the boundaries of clusters consequently become more extreme than those innermost members, as one can see in Fig.~\ref{fig:figure9_3} that $\overline{x_{INP}}(t)$ is terminally larger than $\overline{x_{BP}}(t)$. Cluster-level self-reinforced mechanism forms. As one result, unbiased individuals who can only survive inside the clusters, however, begin to take on positions that are more extreme when interaction strength becomes strong enough to transmit the negative influence from the boundaries of the clusters to the interiors (see Figs.~\ref{fig:figure9_3}(a) and (c)). We can find that the bright speckles become long streaks, and several short streaks start to appear with the same direction in the second and forth quadrants (see Fig.~\ref{fig:figure9_1}(a5)). 

(iv) When the population gets polarized for $A>A_{c2}$, both innermost members and those at the boundaries of positive clusters keep moving towards more extreme positions (see Fig.~\ref{fig:figure9_3}). Moreover, the neighbours of individuals in the clusters of size 1 and the peripheries of modest clusters are evenly split according to the signs of their opinions. Consequently, in addition to more smaller modest clusters (see Figs.~\ref{fig:figure9_2}(a4) and (a6)) and the shrinking largest clusters of opposing camps (see Fig.~\ref{fig:figure9_2}(a2)), smallest clusters of size 1 are more abundant (see Figs.~\ref{fig:figure9_2}(a3)). They guarantee the strongest self-reinforced mechanism. We have longer speckles through the origin of coordinates (see Fig.~\ref{fig:figure9_1}(a6)), and those streaks which stretch out to the first and third quadrants (see Fig.~\ref{fig:figure9_1}(a6)).

The results from statistic networks makes the physical interpretation about the origin of HD state more clear. Unlike time-varying networks, cluster-level self-reinforced mechanism can be present in static social networks because of fixed connections. We firstly find that the abundance of clusters of size 1 and modest clusters may actually be more important than their sizes in determining the effects of cluster-level self-reinforced mechanism, and guaranteeing a balance between repulsive and attractive forces. 
Secondly, one can avoid extreme polarization by increasing $T$ and $\alpha$, but sustaining intermediate $A$. 
Under such parameter condition, people who are reluctant to hearing different opinions can create many different relatively modest ideological islands that people will be stuck on. 
This may leads to low exposure which may prevent the emergence of two opposing extreme opinions~\cite{Axelrod2021}. 
Consequently, extreme polarization is replaced by high-level polarization (compare the results reported in Figs.~\ref{fig:figure3},  \ref{fig:figure4} and Fig.~\ref{fig:figure8}).
When cluster-level self-reinforced mechanism is present, like-minded group members persistently instead support those on the boundaries to repel distant ones, although the innermost members are relatively moderate (see Fig.~\ref{fig:figure9_3}). 
Polarized views can more easily arise from the interaction between individuals at the boundaries of clusters (see Fig.~\ref{fig:figure9_3}), making polarization more likely. One can thus find larger red regions in Fig.~\ref{fig:figure8} than in Figs.~\ref{fig:figure3} and \ref{fig:figure4}, which is associated with lower likelihood of a triple point.

The same is the case for heterogeneous susceptibility, still the cluster-level self-reinforced mechanism is positively related to $A$. 
We can thus observe slight shrinkage of the regions of GC and HD phase when $A\lesssim 2.0$. 
In contrast, regions of P phase instead compress the other two phases with increasing $A$ (see Fig.~\ref{fig:sfigure6}). 
We have confirmed that the mechanism still works, and to achieve HD state is still impossible.

Actually, we have simulated the proposed model on other heterogeneous social networks. Similar heatmaps of SD can still be obtained. Still, the results suggest that evolution outcomes are mainly dominated by the model rules, rather than the structure of employed networks. Note, however, more small denser communities (i.e. more inner connections) can induce stronger effect of cluster-level self-reinforced mechanism, further increasing the likelihood of polarization.

\section{Discussions and Conclusions}
\label{sec:conclusion}
In this paper, we have proposed a simple model based on one core assumption that individuals tend to amplify difference from others with dissimilar opinions according to negative influence, and to attract toward others with similar opinions due to positive influence. 
We correspondingly consider three key realistic ingredients to regulate the opinion dynamics: interaction strength, individuals' susceptibility to opinion dissimilarity between them and their neighbor, and individuals' tolerance to different views determines whether the result of an interaction is attractive or repulsive. 

For time-varying networks, within our model HD state emerges for sufficient susceptibility, intermediate interaction strength and high tolerance, due to a balance between repulsive and attractive forces. The balance is dynamic, which can result in a fluctuated behavior of HD opinion cluster in spectrum space. Especially, the interaction strength determines the ranges at which the balance is disrupted.  
Strong interaction or low tolerance makes HD state not accessible, and the population instead gets polarized. 
Remarkably, the simple model rules successfully generate the three phases, along with three different transitions. 
As another novel finding, there exists a triple point in planes of $(A,~T)$ or $(\alpha,~T)$. 
To the best of our knowledge, these two important and interesting phenomena have never been uncovered and mentioned in previous studies, especially those merely considering reinforcement or influence mechanism.  

In the second part of the paper, we have extended our study to the case where individuals' susceptibility or tolerance follows a power-law distribution whose exponents are $\eta$ and $\xi$, respectively. It is evident that the effects of heterogeneous individuals' intrinsic attributes such as susceptibility or tolerance cannot be ignored. 
We find that the effect of susceptibility heterogeneity is less pronounced, however, it can inhibit HD state to a certain extent, and instead promote polarization.
Likewise, heterogeneous tolerance makes polarization inevitable. 
A global neutral consensus is nearly unaccessible unless both $A$ and $\alpha$ are rather small.  
Out of the three realistic ingredients, interaction strength is particularly prominent as it largely determines the population's outcome in such cases.

At last, we have applied our model on empirical static networks where the interactions among individuals are fixed. 
Although the phase layouts are similar to the case of time-varying networks, the population has larger regions of P phase where extreme polarization coexist with high-level polarization. 
We have uncovered that cluster-level self-reinforced mechanism is responsible for these phenomena, which is dependent on whether spatial  clusters with the same opinion signs are abundant. 
In presence of such mechanism, the struggles among clusters formed by like-minded individuals protects inner ones from the influence of the majority of the population, allowing them to persistently support the group members on the boundaries in face of distant individuals of other clusters. 
It is also the cause of the other result: the triple point is less likely to obtain.    
Moreover, we have confirmed that, for heterogeneous susceptibility or tolerance, this mechanism still works, and to achieve a HD state thus becomes harder or even impossible.

For the first time, this paper proposes a basic theory framework to understand the formation of HD, and we can conclude the five following remarks from this paper:
(i) Most importantly, through simplest possible setting, our proposed model can generally generate three phases: GC, HD and P, as well as triple points, regardless of that the interactions are time-varying or fixed. Which has never been stressed or mentioned by previous opinion models.  
(ii) We have developed an effective method to identify the boundaries between different phases through calculating the maximum susceptibility of opinion entropy, which is confirmed by numerical simulations, allowing us to build phase diagrams and to locate where the triple points are.
(iii) Heterogeneous susceptibility and tolerance turns out to be an inhibiting factor for achieving HD state, which should be avoided. 
(iv) Fixed interactions create a negative impact on the emergence of HD through cluster-level self-reinforced mechanism induced by abundant clusters of size 1 and modest opinion clusters, which can unexpectedly promote polarization.
(v) In addition, the present simple model rules successfully generate the fluctuated behaviors of HD opinion cluster in spectrum, without the noise effect which has been previously considered as the central mechanism for the fluctuation of opinion clusters~\cite{Mas2010,Noorazar2020}.  

It is worth noticing that our model is based on a minimal number of assumptions. 
It does not take into account some empirical features of networks or individuals which might generate different scenarios, such as heterogeneous duration time of interactions, and or different social positions of individuals.
In light of this fact, it is essential to further extend our model framework in the future studies focusing on the existence of societal coherence and integration of plurality and diversity~\cite{Renn2022}. 
Also, this study opens one interesting issues for future research: whether HD state becomes more easily accessible with the involvement of some optimization strategies? And if the answer is yes, how much it will do.

\section*{Acknowledgments}
This work was supported by the Key Program of the National Natural Science Foundation of China (Grant No.~71731002), and by Guangdong Basic and Applied Basic Research Foundation (Grant No.~2021A1515011975). P.-B. C. thanks Kai Qi for helpful discussions.

\section*{Appendices}
\label{sec:appendix}
\begin{figure*}[!ht]
\centering
\includegraphics[width=\linewidth]{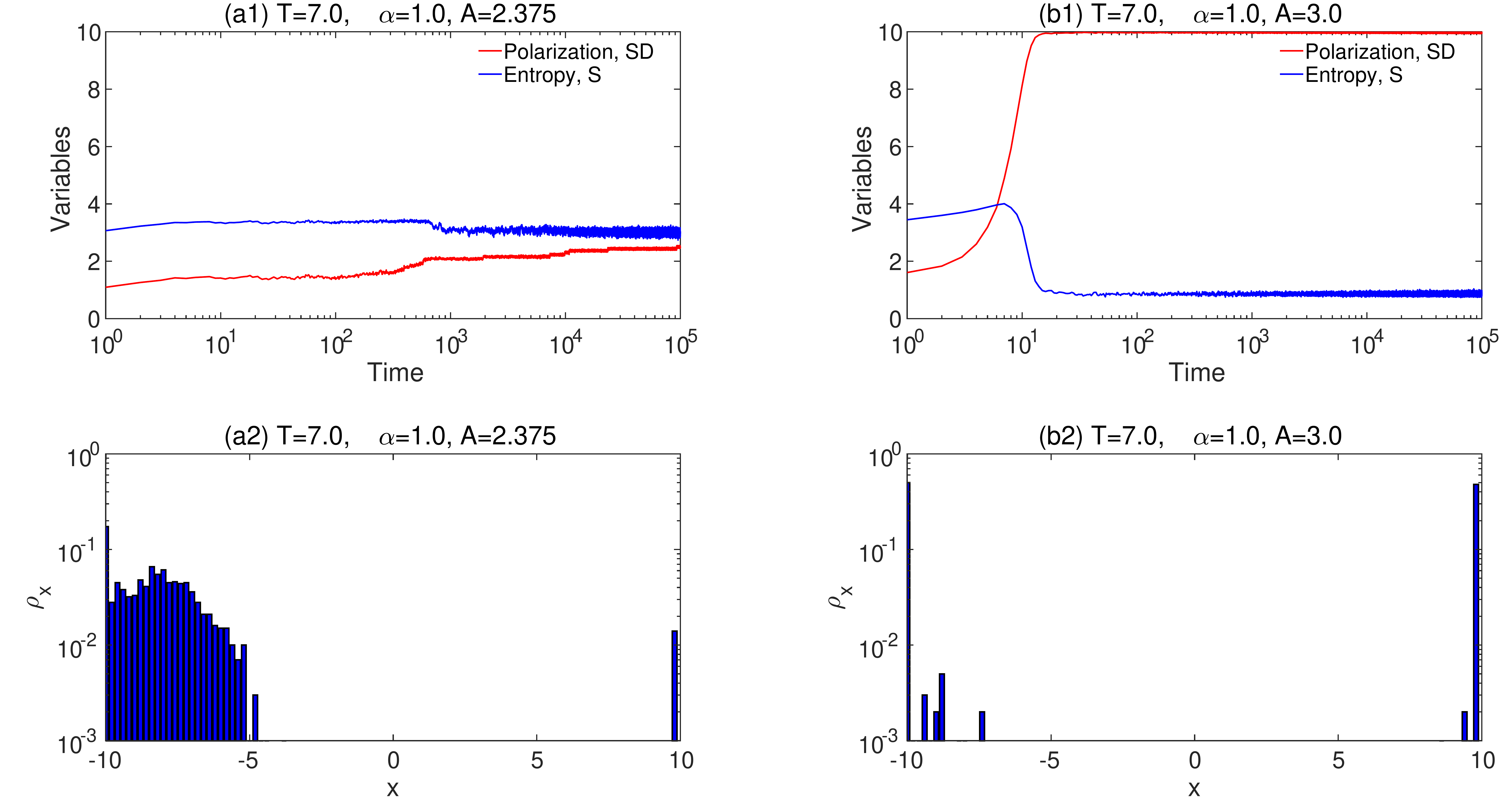}
\caption{(a1)(a2)The evolution of opinion entropy and SD for two different values of $A$. (b1)(b2) The opinion distributions for two different values of $A$. The values of parameters are listed in the titles of the subfigures.
}
\label{fig:sfigure0}
\end{figure*}

\begin{figure*}[!ht]
\centering
\includegraphics[width=\linewidth]{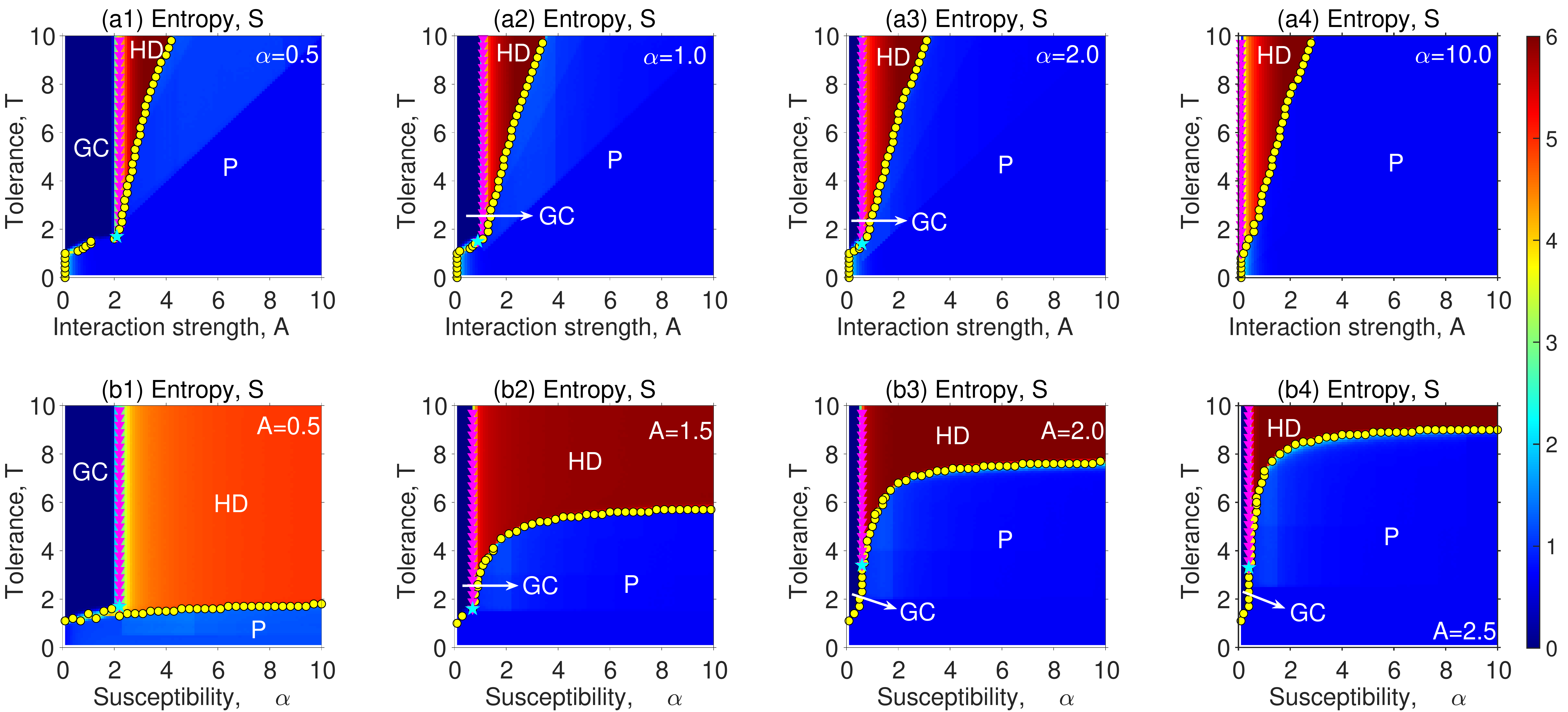}
\caption{(a1)-(a4) The dependence of opinion entropy $S$ on $A$ and $T$ for four different values of $\alpha$. (b1)-(b4) The dependence $S$ on $\alpha$ and $T$ for four different values of $A$. Different regions of the three states are correspondingly labeled. The lines consisting of different markers denote the boundaries between different phases, which are the same as those presented in Fig.~\ref{fig:figure3}. The light blue pentagrams indicate the triple points.
}
\label{fig:sfigure1}
\end{figure*}

\begin{figure*}
\centering
\includegraphics[width=\linewidth]{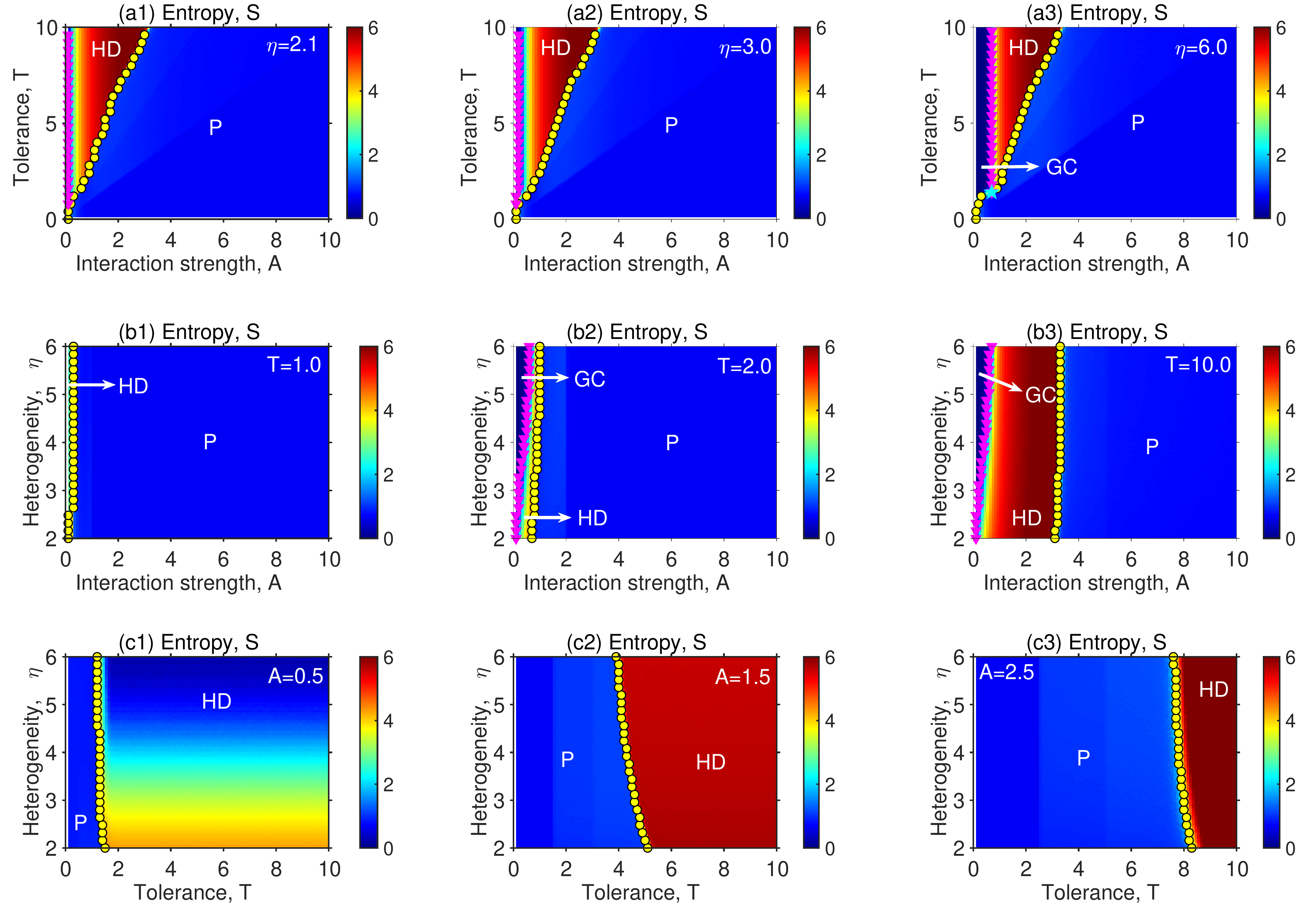}
\caption{(a1)-(a3) The dependence of $S$ on $A$ and $T$ for three different values of $\eta$. (b1)-(b3) The dependence of $S$ on $A$ and $\eta$ for three different values of $T$. (c1)-(c3) The dependence of $S$ on $T$ and $\eta$ for three different values of $A$. Different regions of the three phases are correspondingly labeled. The lines consisting of different markers denote the boundaries between different phases, which are the same as those presented in Fig.~\ref{fig:figure3}. There is only one triple point indicated by light blue pentagram in (a3).
}
\label{fig:sfigure2}
\end{figure*}

\begin{figure*}
\centering
\includegraphics[width=\linewidth]{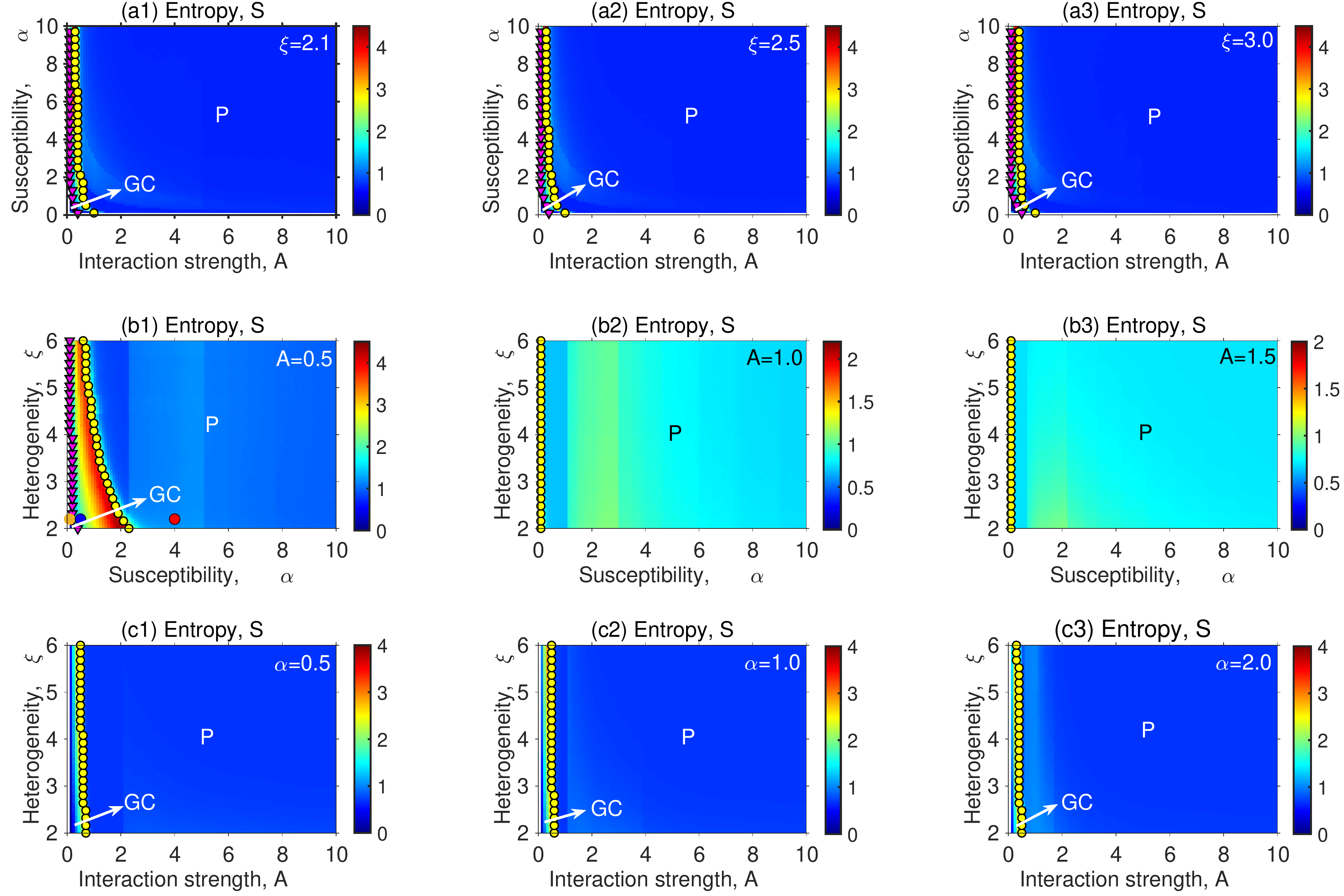}
\caption{(a1)-(a3) The dependence of $S$ on $A$ and $\alpha$ for three different values of $\xi$. (b1)-(b3) The dependence of $S$ on $\alpha$ and $\xi$ for three different values of $A$. (c1)-(c3) The dependence of $S$ on $A$ and $\xi$ for three different values of $\alpha$. Different regions of the two states GC and P phase are correspondingly labeled. 
}
\label{fig:sfigure3}
\end{figure*}

\begin{figure*}
\centering
\includegraphics[width=\linewidth]{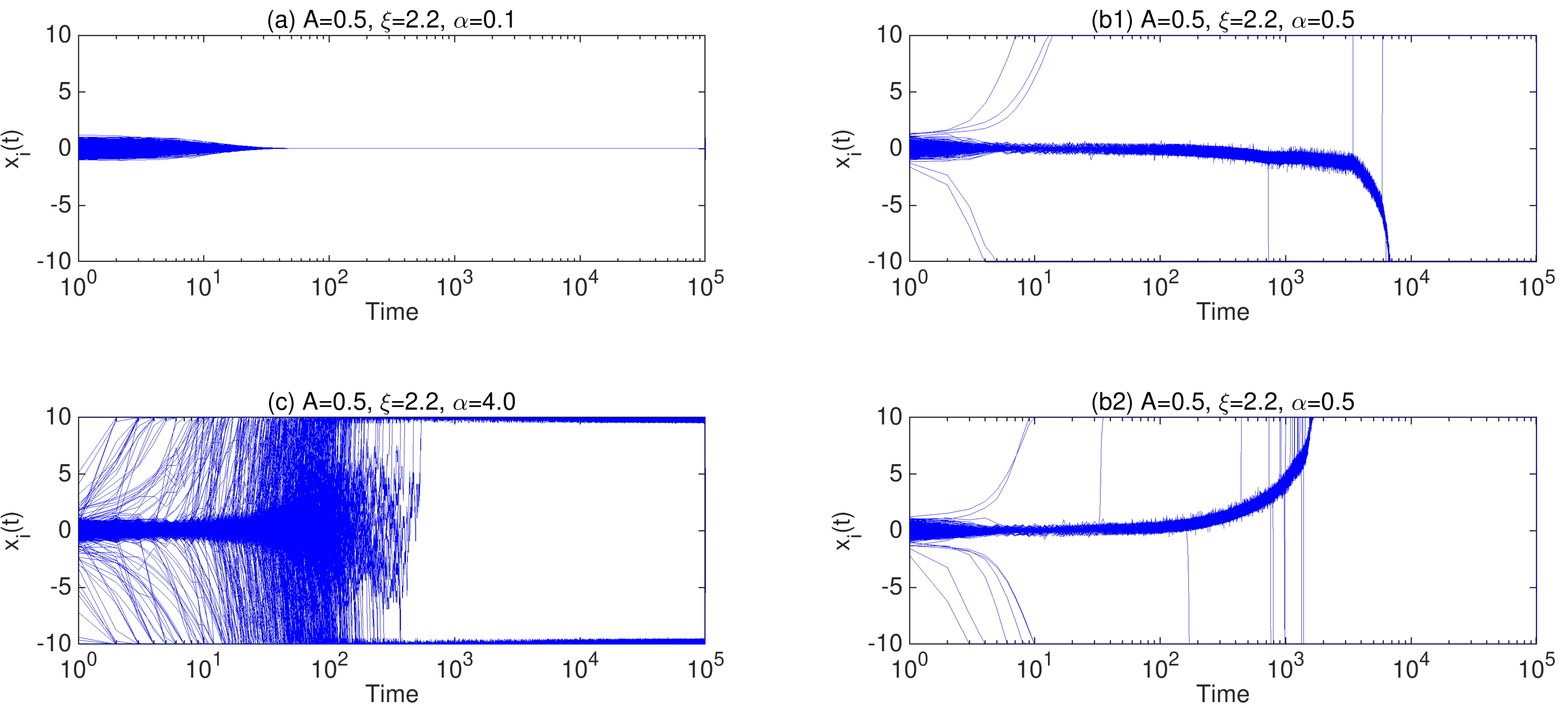}
\caption{The evolution of the entropy of opinion distribution for three different values of susceptibility $\alpha$. The values of other parameters such as interaction strength $A$ and tolerance threshold exponent $\xi$ are listed in the titles of subplots.
}
\label{fig:sfigure4}
\end{figure*}

\begin{figure*}
\centering
\includegraphics[width=\linewidth]{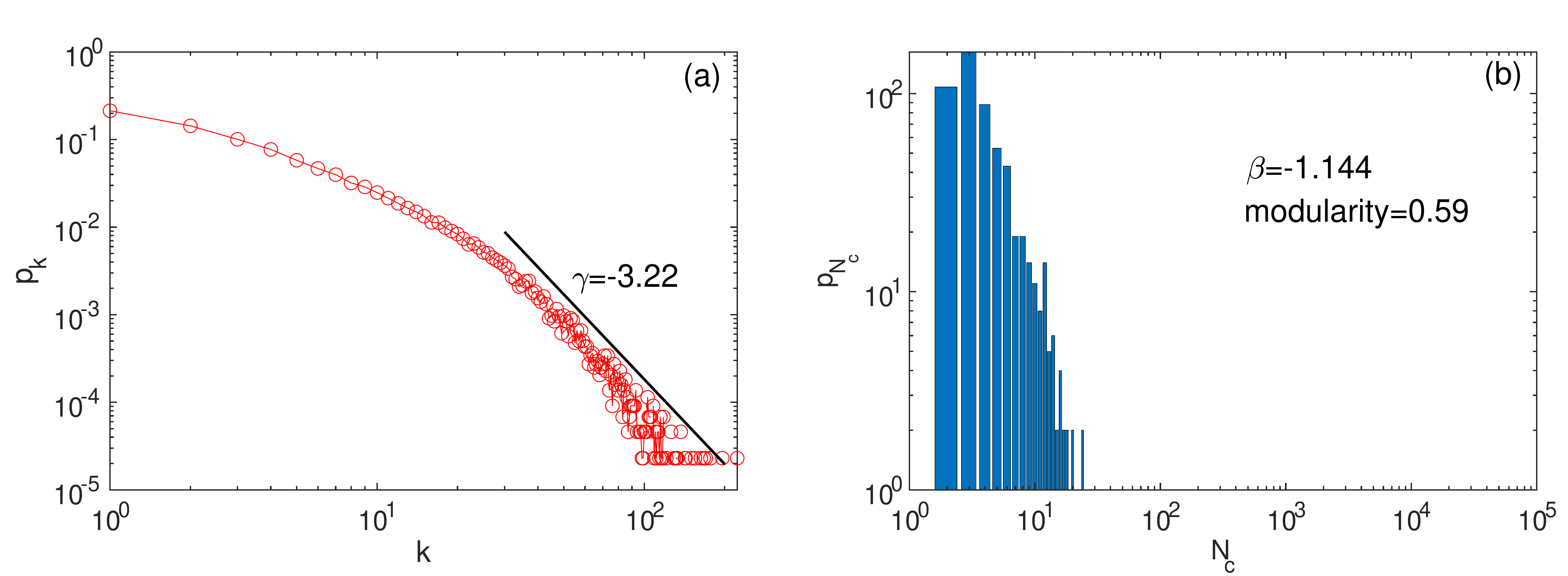}
\caption{The structure details of the part of Facebook network. (a) Degree distribution of the network, along with the fitted degree exponent of $p_{k}\sim k^{\gamma}$. (b) The community size distribution, where the values of modularity and heterogeneity ($p_{N_{c}}\sim N_{c}^{\beta}$) are presented, respectively. The size of the largest community is $14286$. The other structural parameters are: $N=43952$, mean degree $\langle k\rangle =8.30$, $\langle k^{2}\rangle =205.36$, maximum degree $k_{max}=223$, minimum degree $k_{min}=1$, clustering coefficient $c=0.12$ and degree correlation coefficient $r\gtrsim 0$.
}
\label{fig:sfigure5_1}
\end{figure*}

\begin{figure*}
\centering
\includegraphics[width=\linewidth]{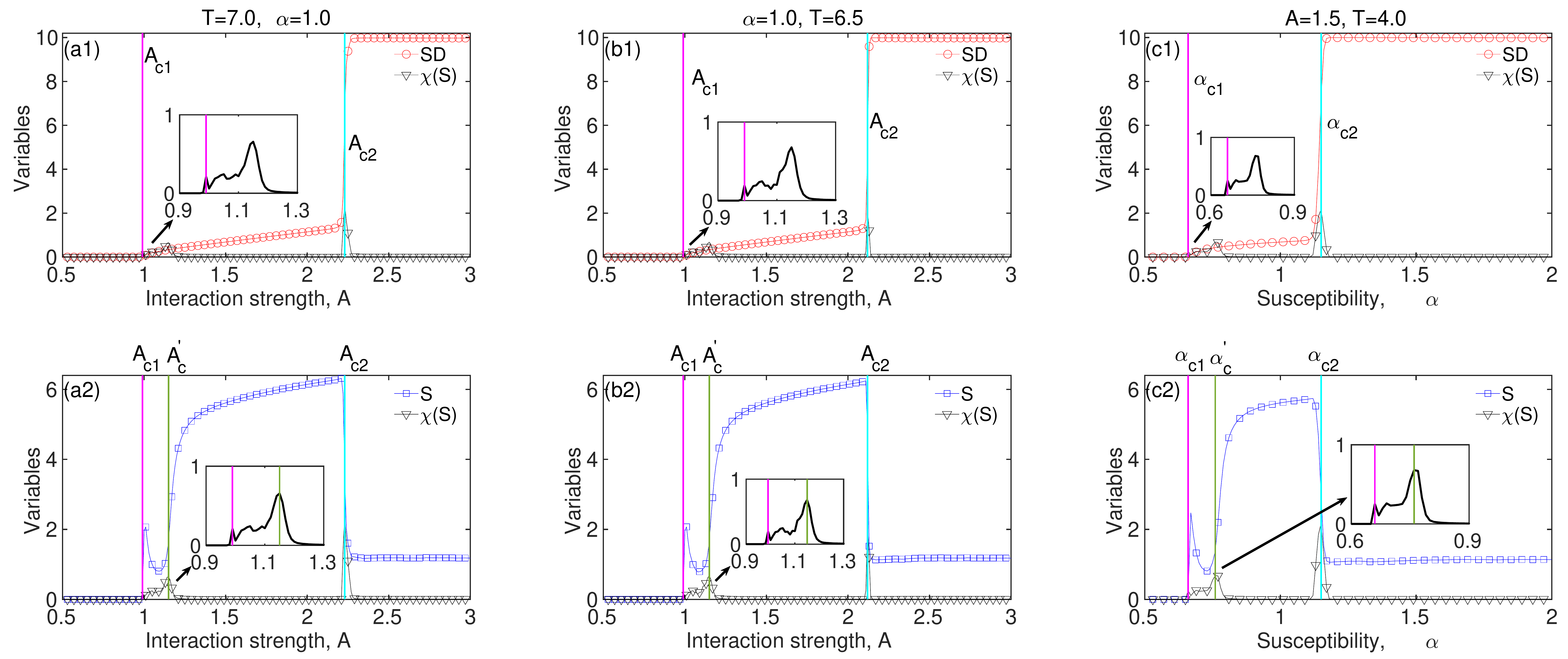}
\caption{The dependence of $SD$ (red circles), $S$ (blue squares), $\chi(S)$ (black curves in the insets) on $A$ in part of Facebook network. Pink vertical line labels the position of $A_{c1}$ at which HD state begins to appear, while light blue vertical line indicates the position of $A_{c2}$ at which HD state vanishes. In particular, $A^{'}_{c}$ denotes an inflection point in $S$ followed by a sharp growth of $S$. The values of $T$ and $\alpha$ are correspondingly listed in titles of (a1), (b1) and (c1). 
}
\label{fig:sfigure5}
\end{figure*}

\begin{figure*}
\centering
\includegraphics[width=\linewidth]{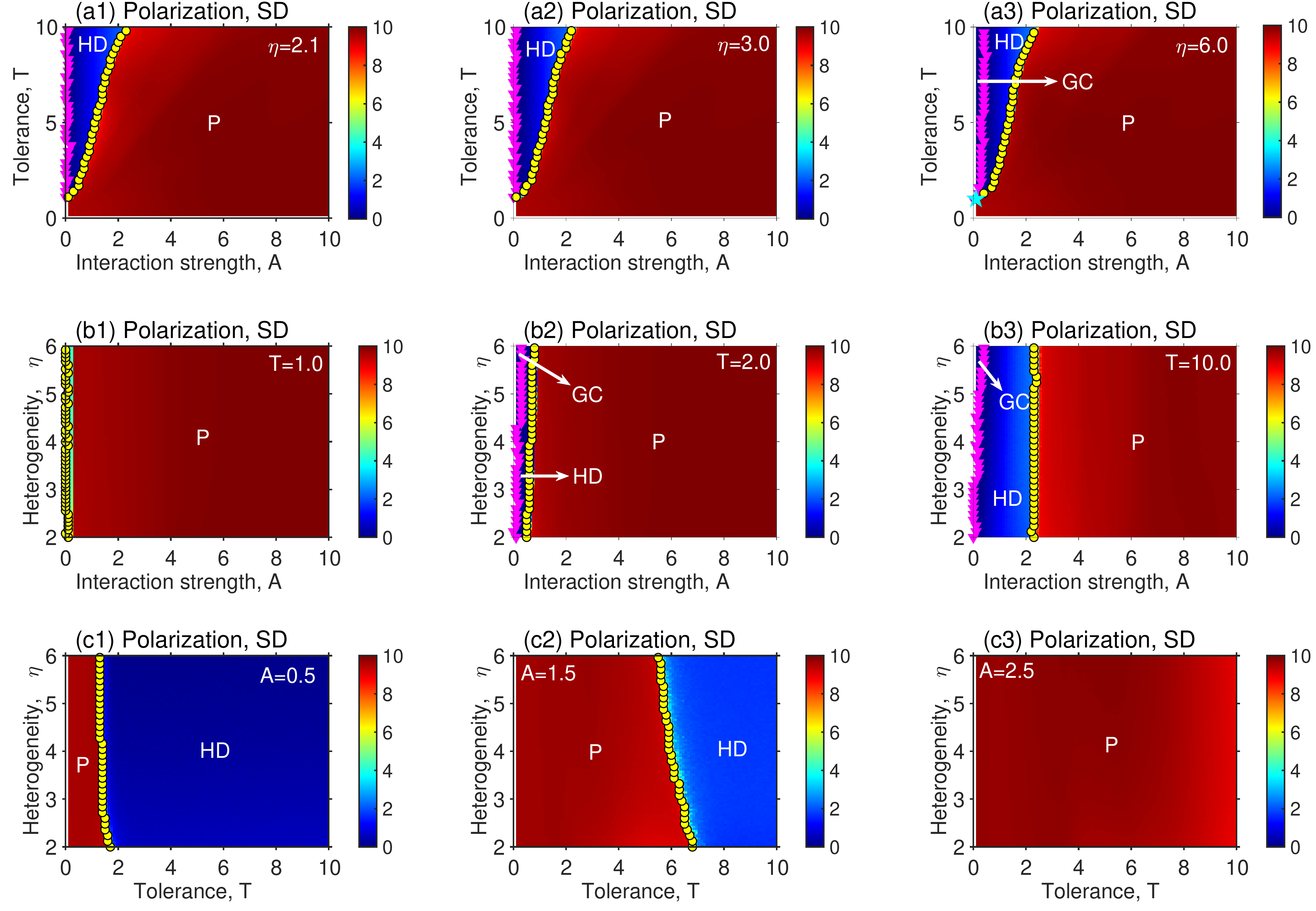}
\caption{Phase diagrams for the population embed in part of Facebook network, where interaction among individuals are fixed. In more detail, (a1)-(a3) phase diagrams in $(A,~T)$ space for three different values of $\eta$. (b1)-(b3) phase diagrams in $(\alpha,~\eta)$ space for three different values of $T$. (c1)-(c3) phase diagrams in $(T,~\eta)$ for three different values of $A$. We run simulations with polarization degree $SD$ in all subplots. The lines consisting of different markers denote the boundaries between different phases, which are the same as those presented in Fig.~\ref{fig:figure3}. The regions belonging to the three phases are correspondingly labeled in the subplots. The light blue pentagram presented in (a3) indicates the triple point in $(A,~T)$ space. 
}
\label{fig:sfigure6}
\end{figure*}

Appendix section lists Figs.~\ref{fig:sfigure0} to \ref{fig:sfigure6} as supplements for the arguments in the body of the paper.

It can be observed in Fig.~\ref{fig:sfigure0} that temporal trajectories of $S$ and $SD$ for $A=2.375$ is definitely different from those of extreme polarization ($A=3.0$). On the other hand, since $A=2.375$ is rather close to $A_{c2}$, it becomes rather hard to keep a balance between repulsive and attractive forces. The repelled force rising from strong negative influence can occasionally outperform attractive force, leading to sharp fluctuations in terms of $S$ (see the peaks indicated by bright blue vertical lines in Figs.~\ref{fig:figure1}(a1) and (a2)). It is thus not weird that the system is possibly asymmetrically polarized for some realizations even if $A\lesssim A_{c2}$, as shown in Figs.~\ref{fig:sfigure0}(a1) and (a2). In such case, the fluctuated integrated opinion cluster splits into two separated ones of different size. Whatever, a main opinion cluster can still be sustained in opinion space (see Fig.~\ref{fig:sfigure0}(a2)). As $A$ getting larger, extreme polarization with nearly equal opposed camps begins to instead emerge (see Figs.~\ref{fig:sfigure0}(b1) and (b2)).

\bibliography{multireference}
\end{document}